 \documentclass[11pt,a4paper]{article}
\pdfoutput=1 % if your are submitting a pdflatex (i.e. if you have images in pdf, png or jpg format)
\usepackage{jcappub}	% jheppub includes hyperref,color, natbib, amsmath, amssymb, epsfig, graphicx
 \usepackage{graphicx,amssymb,amsmath,amsfonts}
\usepackage{hyperref}
 \usepackage{diagbox}
\usepackage{makecell}
\usepackage{latexsym}
  \usepackage{tikz}
\usetikzlibrary{positioning,arrows}
\usetikzlibrary{decorations.pathmorphing}
\usetikzlibrary{decorations.markings}
\usetikzlibrary{snakes}
\usetikzlibrary{matrix}
\usepackage{slashed}

\usepackage{graphicx}
 
\numberwithin{equation}{section}

\usepackage[utf8]{inputenc}

\date{\today}

\def\be{\begin{equation}}
\def\ee{\end{equation}}

\newmuskip\pFqmuskip

\newcommand*\pFq[6][8]{%
  \begingroup % only local assignments
  \pFqmuskip=#1mu\relax
  \mathchardef\normalcomma=\mathcode`,
  \mathcode`\,=\string"8000
  \begingroup\lccode`\~=`\,
  \lowercase{\endgroup\let~}\pFqcomma
  {}_{#2}F_{#3}{\left(\genfrac..{0pt}{}{#4}{#5}\Big | #6\right)}%
  \endgroup
}
\newcommand{\pFqcomma}{{\normalcomma}\mskip\pFqmuskip}
\newcommand{\pmat}{\begin{pmatrix}}
\newcommand{\fpmat}{\end{pmatrix}}
\newcommand{\eq}{\begin{equation}}
\newcommand{\feq}{\end{equation}}
\newcommand{\cas}{\begin{cases}}
\newcommand{\fcas}{\end{cases}}

\newcommand{\eqarray}{\begin{eqnarray}}
\newcommand{\feqarray}{\end{eqnarray}}

\newcommand{\e}{\eta}

\newcommand{\f}{\phi}

\newcommand{\g}{\gamma}

\def\a{\alpha}				\def\g{\gamma}		
\def\e{\varepsilon}

					\def\f{\phi}

						\def\L{\Lambda}

\def\be{\begin{equation}}
\def\ee{\end{equation}}
\def\bea{\begin{eqnarray}}
\def\eea{\end{eqnarray}}
\newcommand{\nn}{\nonumber}
\newcommand{\ft}[2]{{\textstyle\frac{#1}{#2}}}

\title{Integrable models of inflation \\ beyond slow-roll}

\author[a,b]{M. Bianchi}
\author[a,b]{, G. Dibitetto}
\author[b]{, J. F. Morales}
\author[a]{and V. Zevola}
\affiliation[a]{ Dipartimento di Fisica, Universit\`a di Roma ``Tor Vergata", Via della Ricerca
Scientifica 1, 00133, Roma, Italy.}
\affiliation[b]{INFN section of Rome ``Tor Vergata", Via della Ricerca
Scientifica 1, 00133, Roma, Italy.}

\emailAdd{bianchi@roma2.infn.it, dibitetto@roma2.infn.it, morales@roma2.infn.it, vincenzozevola@outlook.com}

\abstract{
We propose a novel analytic approach to the study of multi-component FLRW cosmologies and their perturbations. 
{The dynamics is triggered by a single scalar field with a scalar potential codifying the energy density and pressure of the multi-component fluid.} 
 This description unifies standard Big Bang cosmologies, models of inflation and dark energy under a unique framework. The key to integrability is to express the scalar potential { in terms 
 of the Hubble function $H(\phi)$ that plays the role of a fake superpotential, turning the dynamics into a first order problem, that may be analytically solved in a suitable time coordinate}.
In this framework, we propose integrable inflationary models with similar properties to the ones analysed in the literature and compatible with observations. 
Finally, we study scalar and tensor cosmological perturbations in each model by {integrating Mukhanov-Sasaki equations  via} numerical  and (semi-)analytic techniques. This allow us to compute the power spectrum, the spectral indices and the tensor-to-scalar ratio beyond the slow-roll approximation and compare our general results {against} the currently available observations and the theoretical predictions {based on} the slow-roll approximation.

     }

\begin{document}

\tikzset{
line/.style={thick, decorate, draw=black,}
 }

\maketitle

 \section{Introduction}
\label{section:intro}
The inflationary paradigm was proposed at the beginning of the 1980s \cite{Starobinsky:1980te,Guth:1980zm,Linde:1981mu,Linde:1983gd,Linde:1986fd} as a possible solution to several puzzles that arise in the context of the standard Big Bang cosmology\footnote{For recent reviews, see {\it e.g.} \cite{Riotto:2002yw}. For an effective field theory approach see \cite{Cheung:2007st}. For extra natural Inflation see \cite{Arkani-Hamed:2003xts}.}. More specifically, Big Bang cosmologies are affected by the so-called homogeneity, isotropy, horizon and flatness problems, which can be overcome in this context by means of phase of accelerated expansion \emph{a.k.a.} cosmic inflation.
Over the last four decades, many different concrete inflationary models have been proposed, arising from the coupling between Einstein gravity and one or more scalar field(s). These models provide viable mechanisms to drive the universe from its origin to the large-scale structure formation. 

However, despite its wide success at a theoretical level, it is incredibly hard to probe inflation through cosmological observations, simply because it involves very high energy scales. One of the key elements that may allow us to tell a good model of inflation from a bad one is the study of cosmological perturbations predicted by the given model. An important class of such tiny primordial fluctuations has grown to observable size thanks to inflation and ever since the time of recombination has frozen out and is still detectable nowadays by carefully analyzing the CMB spectrum.

Generically, the so-called slow-roll inflationary models tend to predict a nearly Gaussian power spectrum for primordial perturbations, as well as a small tensor-to-scalar ratio. These generic features are perfectly compatible with the current status of cosmological observations coming from the CMB data. Still, properly compelling evidence for the inflationary paradigm is yet to be supported by observations. The increasing precision of the CMB measurements from COBE \cite{COBE:1992syq}, to WMAP \cite{WMAP:2012fli} and PLANCK \cite{Planck:2018vyg} definitely motivates a better theoretical control on the various inflationary proposals when it comes to their predictions for the power spectrum of perturbations. We refer, \emph{e.g.} to \cite{Kallosh:2025ijd} for a wide panoramic view on the current status of the different inflationary models in the light of the most recent observational develpoments. Phenomenologically interesting proposals are certainly the one of the so-called Higgs inflation \cite{Salopek:1988qh,Bezrukov:2007ep} (based on the Higgs field of the SM as the inflaton), and the class of conformal attractors, which is recently proposed in an even more general framework in  \cite{Kallosh:2026qrc}.

Our present goal in this context is to produce novel models of inflation for which one can make use of analytic tools to gain better control on the parameter space of a given model in connection to observable predictions. 
%To this end, we will start by introducing a tool to systematically generate \emph{integrable} inflationary models that obey first order dynamics. This tool is based on formulating the classical field equations in terms  of a so-called fake superpotential \cite{SkendTownsEtc}, borrowed from the language of supergravity and holography, where it found many applications \cite{SkendEtal} even in cosmological contexts \cite{SkendHoloCosmo} with different goals from our present ones. Subsequently, we will make use of this language to propose a few novel classes of single field inflationary models that enjoy similar features to other popular models. 
To this end we consider Friedmann-Lemaitre-Robertson-Walker expanding universes where evolution is triggered by a scalar field minimally coupled to gravity with scalar potential codifying a multi-component fluid dynamics. The key observation is that Friedman 
equations relate the scalar potential $V(\phi)$ to the Hubble function $H(\phi)$, that plays the role of a fake superpotential, reducing Einstein equation into a first order system that may be analytically solved  in a suitable time coordinate.
 In this framework, we construct  integrable models of inflation obtained as deformations of standar Starobinsky, alpha-attractors, polynomial, hyperbolic and trigonometric inflation potentials, and multi-component fluid universes describing the later evolution. 

In each inflationary model, we then analyze the spectrum of cosmological perturbations by semi-analytical techniques adapted from \cite{Bianchi:2024mlq}. 
Using the universe scale $a$ as time coordinate, we integrate Mukhanov-Sasaki  equations for both scalar and tensor perturbations via numerical and semi-analytic techniques, compute the power spectra  $P_S(k)$ and $P_T(k)$,  their spectral indices $n_s$ and $n_T$ and the tensor-to-scalar ratio $P_T(k)/P_S(k)$ beyond the `standard' slow-roll approximation. Finally we compare our general results with the currently available observations and the theoretical predictions in the slow-roll approximation.

The plan of the paper is the following. In Section {2} we describe our approach in great details and discuss how to obtain integrable cosmologies from scalar dynamics relying on the `Hubble function'. We then compare different time coordinates and for illustration discuss 
cosmological models driven by a mono-component fluid and a toy-model for a multi-component fluid  perfectly mimicking the very successful $\Lambda$CDM model.
 
Several classes of integrable models of inflation are analyzed in Section {3}. we start with the integrable version of the celebrated Starobinsky model, and then pass to consider (integrable versions of) polynomial inflation, $\alpha$-attractors, trigonometric models, and  hyperbolic models. 
Section {4} is devoted to study of cosmological perturbations by means of semi-analytic tools that allow to go beyond the slow-roll approximation. In Section {5}, we then extract the relevant observables {\it i.e.} the power spectra, the spectral indices and the tensor-to-scalar ratios and compare our results with the ones that can be obtained in the slow-roll approximation.

We conclude in Section {6} with the summary of our results and an outlook on possible generalizations and (phenomenological) applications.

 \section{A novel approach: integrable cosmologies from scalar dynamics}
\label{section:novel}
   The cosmic evolution of a homogenous and isotropic universe is governed by the Friedmann equations, relating the Einstein tensor for  a Friedmann-Lemaitre-Robertson-Walker (FLRW) metric\footnote{For simplicity, we take flat spatial sections ($\kappa=0$, {\it i.e.} $\Omega = 1$), as favoured by all available cosmological observations.}
  \be
  ds_4^2=-dt^2+a(t)^2 d{\bf x}^2 \ ,\label{flrw}
  \ee 
   to the stress energy tensor of a (generically non-perfect) fluid, which is completely specified in its locally inertial frame,  by its energy density $\rho(t)$ and its pressure $p(t)$.  In this paper, we  propose to model the fluid dynamics by introducing an auxiliary self-interacting scalar field $\phi$, with potential
   $V(\phi)$, minimally coupled to Einstein gravity,  through the action
\be
S=\int d^4x \sqrt{-g} \left[\ft{M_{\rm pl}^2}{2} R -\ft12 (\partial \phi)^2-V(\phi)   \right] \ .\label{sphi}
\ee
with\footnote{ Strictly speaking, except for the initial conditions for inflation, that we will take to be set by the so-called Bunch-Davies vacuum, our analysis is entirely classical. So the $\hbar$ in the `standard' definition(s) $M_{\rm Pl}^2 = \hbar c /8\pi G_N$  ($\ell_{\rm Pl} = \hbar/ M_{\rm Pl}c$) is actually absent  in the Einstein-Hilbert action.}  $M_{\rm pl}^2=(8\pi G_N)^{-1}$.

\subsection{The fake superpotential}
The field equations following from the action (\ref{sphi}) read
\bea
 R_{MN}{-}\ft12 g_{MN} R& =& M_{\rm pl}^{-2} T_{MN} \ ,\nn\\
  \Box \phi- V'(\phi) &=& 0\ ,
\eea
with  
\be
T_{MN}=    \partial_M \phi \partial_N \phi {-} g_{MN}  \left[ \ft12 (\partial \phi)^2+  V(\phi) \right]\ .
\ee
 In the following we use units where $M_{\rm pl}=1$. 
% \bea
% \rho(a) ={ -}T^0_0 &=& ~~ V(a)+{a^2 H(a)^2\over 2}  \phi'^2=3 m_{\rm pl}^2 H(a)^2  \nn\\
%  p(a) =~~ T^i_i  &=&{-}V(a)+{a^2 H(a)^2\over 2} \phi'^2=-m_{\rm pl}^2 \left[2 a H(a) H'(a)+3 H(a)^2\right]
% \eea
We  consider spatially flat expanding universes, and exploit reparametrization invariance to rewrite the metric (\ref{flrw})  in a new time variable $\tau(t)$. As a consequence, we start from the \emph{Ansatz}
\bea
\label{FLRW_1}
ds_4^2 \, &=&\, -c(\tau)^2 d\tau^2 + a(\tau)^2 \, ds_{\mathbb{R}^3}^2 \ ,   \nn\\
\phi &=& \phi(\tau)\ .
\eea
In these coordinates, the field equations read
 \bea
&&  a\,  c' \, \phi'  - 3 \,c\, a' \, \phi' -a \,c \,\phi'' -a\, c^3\, \partial_\phi V(\phi
   ) = 0\ ,\nn\\
&&   6 \, a'{}^{2}\,  - 2 c^2\,a^2\, V(\phi  )- a^2\, (\phi')^2 = 0 \ ,\label{einseq} \\
&&  2 \left(2\, a \,  a' \, c'{-}c a'{}^{2}  {-}  2\, c\, a\,   a''  \right) {-}c\, a^2  \phi'{}^{2}{+}2 \,a^2\, c^3\, V(\phi)= 0\ ,\nn
 \eea
  where primes denote derivatives with respect to $\tau$. One may easily check that the first equation follows from the last two, which
  can be written in the form
  \bea
 V&=&  \frac{ 1}{a^2\, c^3}  \left(a\, c\, a''-a \, a'\, c'+2 c a'{}^2\right) \ ,\nn\\
\phi'{}^2 &=&  \frac{2  }{a^2 c} \left(-a\, c\, a''+a \, a'\, c'+c \, a'{}^2\right) . \label{einst2}
    \eea
If we go back and adopt cosmic time as time coordinate (\emph{i.e.} choose $\tau=t$), these equations reduce to  the well-known Friedman equations
\bea
&&     \,\ddot{\phi} + 3 \, H \, \dot{\phi} +  \partial_\phi V(\phi) =0  \ ,\nn\\
&&  \, 3 H^{2}  -  \, V(\phi  )- \, \ft12 \, \dot{\phi}^2 = 0 \ . \label{fleq}
 \eea   
   To solve these equations, or in general (\ref{einseq}), we introduce a function $W(\phi)$ \emph{a.k.a.} fake superpotential, implicitly defined  in terms of the potential $V(\phi)$ by the differential relation
     \be
 V(\phi)=\ft{3}{4 }   W(\phi)^2 -\ft12 W_\phi(\phi)^2   \ . \label{vw}
 \ee
Referring to  $W(\phi)$ as the fake superpotential originates from the fact that, for those models that admit a supergravity embedding, a particular solution to \eqref{vw} is given by the superpotential of theory. In that case, \eqref{vw} may be then interpreted as a Ward identity for supersymmetry. In terms of this $W$, the field equations (\ref{einst2})  reduce to
  \be
{   a'(\tau)  \over   a(\tau)  }= - {  \phi' W(\phi)  \over 2    W_\phi(\phi) } \qquad , \qquad    c(\tau)  = \frac{dt}{d\tau} =-{ \epsilon \phi'(\tau) \over W_\phi(\phi) } \qquad ,\label{eqaa0}
 \ee
 with $\epsilon=\pm 1$ being a sign ambiguity. In the following we choose always $\epsilon=+1$, and the signs of $W$, $\phi$, $\eta$ such as\footnote{Note that $\dot{H}<0$ since $\dot{H}= H_\phi \dot{\phi} = - H_\phi W_\phi = - 2 H_\phi^2 <0$.} 
  \be
  \dot{a},\dot{\eta},\dot{\phi}>0  ~, \qquad\qquad  H_\phi<0\ ,
 \ee
 with the dot denoting derivative with respect to cosmic time $t$. In this time variable, the equations of motion (\ref{eqaa0}) take  the simple form
 \be
{   \dot{a}  \over   a  }=  {  W  \over 2} \qquad , \qquad    W_\phi =  -  \dot{\phi}   \ .\label{eqaa0}
 \ee 
  The first equation relates the superpotential $W$ to the Hubble rate
  \be
 H(t)\equiv {   \dot{a}(t)  \over   a(t)  }={W(t)\over 2} \ ,
 \ee
 leading to the on-shell relation
  \be
 V(\phi)=3  H(\phi)^2 -2 H_{\phi}(\phi)^2    \ .\label{vh}
 \ee
This relation suggests that what can be termed the `Hubble function' $H(\phi)= H(a(\phi))$ is in some sense more fundamental than the potential itself; and the dynamics is better specified by given $H$ rather than $V$.  The crucial simplification is that, using $H$, the Friedman equations reduce to the first order equations in (\ref{eqaa0}), which may be often integrated in an analytical form. 
 
 \subsection{The time coordinate}
 
  The second simplification that plays a crucial role in our analysis is be the choice of time coordinate. 
  In this section, we discuss  different choices of time parametrization and the relations among them.

  \begin{itemize}

  \item{Cosmic time $\tau=t$:
  \bea
  a & =& e^ {  - \int  {H d\phi  \over 2H_\phi }}  ~, \qquad\qquad  V=3 H^2 +  \dot{H}   ~, \qquad\qquad 
 \dot{\phi} = \sqrt{- 2\dot{H} }\ ,
   \nn\\[2mm]
  ds_4^2 &=&  -dt^2+ a^2 d{\bf x}^2\ .
  \eea
  }
  
  \item{Conformal time $\tau=\eta$:
  \bea
  H&=&    {  a_\eta  \over a^2   }~,    \qquad   V =  \frac{  a_{\eta\eta}}{a^3} + \frac{  a_\eta^2 }{a^4}   ~,   \qquad    \phi_\eta   =\sqrt{ -  2 a \, \partial_\eta\left(  {  a_\eta  \over a^2   }\right)}  ~, \quad \dot{\phi}=\sqrt{ -  {2\over  a} \, \partial_\eta\left(  {  a_\eta  \over a^2   }\right)}\ ,  \nn\\
  ds^2 &=&  a^2 (-d\eta^2+ d{\bf x}^2)\ .
  \eea
  }
  
  \item{ Scalar field time $\tau=\phi$  
 \bea
 a(\phi) &=&  e^{  -{1\over 2  } \int  {H d\phi\over H_\phi } }  ~ , \qquad\qquad V = 3 H^2 -2 H_\phi^2 ~ , \qquad\qquad \dot{\phi}=-2 H_\phi \ ,  \nn\\
  ds_4^2 &=&  -{d\phi^2\over 4 H_\phi^2}+a^2 d{\bf x}^2\ .
 \eea
  }
  
   \item{  {Scale factor} time $\tau=a$  
 \bea
  V& =&   3 H^2+a H_a H ~ , \qquad\qquad  \phi_a  = \sqrt{ -\frac{2H_a}{a H} } ~ , \qquad\qquad
 \dot{\phi}= \sqrt{-2 a H_a  H} \ ,  \nn\\
  ds_4^2 &=& - { da^2\over a^2 \, H^2}   + a^2 d{\bf x}^2. \label{metrica}
 \eea
  }
  \end{itemize}
 Here and henceforth subscripts always denote derivatives. The choice $\tau=a$ was widely discussed in \cite{Bianchi:2024mlq}, where FLRW universes sourced by a perfect multifluid are analyzed in detail.
  One can switch from one time variable to the other, using the relations
 \be
 dt=a d\eta=-{d\phi\over 2 H_\phi(\phi)} ={da\over a H(a)} \ . \label{dtime}
 \ee
  The convenience of a given choice of $\tau$ depends on the specific physical situation. 
  We notice that  backgrounds of the type (\ref{FLRW_1}) are isotropic, so they can be interpreted as universes sourced by a fluid with energy density and pressure given by
  \bea
 \rho &=&T_{\mu \nu} u^{\mu} u^{\nu}  =  \frac{ \dot{\phi}^2}{2} + V(\phi)   \ , \nn\\
  p &=& T_{\mu \nu} \left(\ft13 g^{\mu\nu}+u^\mu u^\nu \right)=  \frac{ \dot{\phi}^2}{2} - V(\phi) \ ,  \label{energypressure}
 \eea
 where  $u^\mu u_\mu=-1$.  We will mainly use $a$ as the time coordinate. In these coordinates, the energy density  and pressure  produced by the scalar field respectively read
 \bea
 \rho(a)  &=& 3   H(a)^2 \ , \nn\\
  p(a)  &=& - 2 a H(a) H'(a) -3 H(a)^2 \ . \label{rhopa}
 \eea
    Reversing the logic, given a universe with energy density $\rho(a)$, we can always associate it with a scalar field $\phi$, with the appropriate fake superpotential $W(\phi)$ or, equivalently, a Hubble function $H(a)=\sqrt{\rho(a)/3}$. As a consequence, the FLRW metric (\ref{metrica}) will by construction solve
    the  Friedman-Lemaitre equations (\ref{fleq}) for any choice of $\rho(a)$. 
\subsection{One component universe  }    
   \begin{table}[t]
  \label{tabcomp}
\begin{center}
\begin{tabular}{|c|c|c|c|c|c|}
\hline
 \text{Fluid Type} & Symbol & $ w$ & $n$ &   $\rho$ &$  q $ \\
 \hline\hline
 \text{Vacuum} & $\L$ & $-1$ & $0$   & $1$  &$ -1$  \\
\hline
\text{Strings, Curvature} & $\sigma,\kappa$ & $-\frac{1}{3}$ & 2  & $ a^{-2}   $&$  0 $ \\
\hline
 \text{Matter (dust)} & m & $0$ & $3$   &$ a^{-3} $&$ \ft{1}{2} $ \\
\hline
 \text{Radiation} & $\g$ & $\frac{1}{3}$ & $4$   &$ a^{-4} $&$ 1$ \\
 \hline
\text{Stiff matter}  & $s$ & 1 & 6 &  $a ^{-6}$ & $ 2$ \\
 \hline
\end{tabular}
\end{center}
\caption{\it The salient features of cosmologies for a universe made of a single-component perfect fluid. For each perfect fluid, we display its equation of state parameter $w$, the value of $n$ specifying the scaling of $H$, the scaling of $\rho$ and the corresponding deceleration parameter $q$. We set $3H_0^2\Omega_0 =1$ for simplicity.}
\label{Table:w_fluids}
\end{table}% 
The simplest \emph{Ansatz} that turns out to work for the fake superpotential is a single exponential\footnote{The parameter $n$ is real and positive but not necessarily integer or rational. In many specific cases (radiation, dust, curvature, dark energy, ... stiff matter) it turns out to be an integer.} 
  \be
  H(\phi) =   H_{\rm in}   e^{ -\frac{\sqrt{n}}{2 }\phi} \ ,\label{hphi1}
  \ee 
    leading to the potential
  \be 
   V( \phi )=  H_{\rm in}^2   (3-\ft{n}{2})    \, e^{-\sqrt{n} \phi   } \ .
   \ee
   The equation of motion of the scalar field is then integrated to yield
   \be
    a(\phi) =   e^{  -{1\over 2  } \int  {H(\phi) d\phi\over H_\phi(\phi) } }= e^{{ \phi \over   \sqrt{n} }}\ .
   \ee
   where, for simplicity, we drop the integration constant that can be simply reabsorbed in a rescaling of $a$. 
    
Inverting the above relation, one finds
    \be
   \phi(a) = \sqrt{n} \log(a) \ .
      \ee
    Plugging this into (\ref{hphi1}) one finds
     \be
 H (a)=  H_0  a^{-{n\over 2}}  = \sqrt{\rho(a)\over 3} \label{ha1} \ ,
   \ee
    leading to the metric
   \be
   ds^2= -a^{n-2} \, da^2 +a^2 d{\bf x}^2\ .
   \ee
     Plugging (\ref{ha1})  into (\ref{rhopa}), we may read off that the scalar field effectively behaves  
   as a perfect fluid with equation of state $p=w \rho$ after the identification
   \be
\label{n2w}
   n=3(w+1) \ .
   \ee
      In table \ref{Table:w_fluids} we display  the characteristic parameters for some commonly discussed choices of $w$, along with their physical interpretations.   
   It is worth noticing  that for a single fluid component the time variable $a$ is related to cosmic time $t$ and conformal time $\eta$ via 
  \be
  t(a) = \int {da\over a H(a)} ={2 a^{n\over 2}\over n } \qquad  \ ,   \qquad  
  \eta(a)=   \int {da\over a^2 H(a)}= \frac{2 a^{\frac{n-2}{2}} }{  (n-2)}  \ .
  \ee
  
The general relation for multi-component fluids was spelled out in section 2.2 {\it What's the (cosmological) time?} of \cite{Bianchi:2024mlq}.

 \subsection{Multi component universes and a $\Lambda$CDM model    }
    
A multi-component universe is a universe made of non-interacting species (perfect fluids), with energy density
     \be
    \rho(a)=3 H_0^2 \sum_i \Omega_i a^{-n_i} \ ,
     \ee
      with  $\Omega_i$ specifying the abundances of the given component, $\sum_i \Omega_i=1$ (including the `curvature' component $\rho_\kappa(a)= -\kappa/a^2$),  and $H_0=H(1)$ is the Hubble constant today.  The corresponding Hubble function 
     is given by
      \be
 H(a) =  H_0 \left[ \sum_i \Omega_i a^{-n_i} \right]^{1\over 2} \ .
   \ee
      More generally, one can consider models of cosmological evolution where both the density $\rho(a)$ and the pressure $p(a)$ are rational functions of the universe size $a$. 
      Under these assumptions the equation of state index $w(a)$ is  also a rational function of $a$
       \bea
 w(a) &=& {p(a) \over \rho(a) } =  -1- {2 a H'(a)\over 3 H(a) }\ ,
 % \nn\\
% q(a) &=& -{a \ddot{a} \over \dot{a}^2}   =-1- {a H'(a)\over  H(a) } \label{wqh}
 \eea
 that  can be integrated to find $H(a)$. Indeed given $w(a)$, an arbitrary rational function 
 \be
     \ft{3}{2} \left[ w(a)-1 \right] = \sum_i {\nu_i\over (a-a_i) }\ ,
     \ee
  specified by the poles $a_i$ and the residues $\nu_i$, one finds
     \be
   H(a) =  e^{    \ft{3}{2} \int \left[ w(a)-1 \right] {da\over a}  } =\prod_{i=1}^8 \left( 1-\ft{a_i}{a} \right)^{\nu_i\over a_i}\ .
     \ee
             We notice that the poles $a_i$ and residue $\nu_i$ are in general complex numbers, but the product $H(a)$ is by construction a real function. 
 To illustrate the flexibility of the construction,  let us consider a simple toy model for the universe evolution. We consider a universe evolution specified by the equation of state index
       \be
   w  (a) ={p(a)\over \rho(a)}= \frac{ -\Omega _t\, a^{-8}   + \Omega _s a^{-6}  +\ft13  \Omega _{\gamma } a^{-4} +0 \times \Omega _m \, a^{-3} - \Omega _{\Lambda } }{\Omega _t\, a^{-8} + \Omega _s\, a^{-6} +
    \Omega _{\gamma } a^{-4}  + \Omega _m \, a^{-3} + \Omega _{\Lambda } } \ .\label{warho}
   \ee 

      \begin{figure}[t]
  \begin{minipage}{\textwidth}
    \centering  
\includegraphics[width=0.5\textwidth]{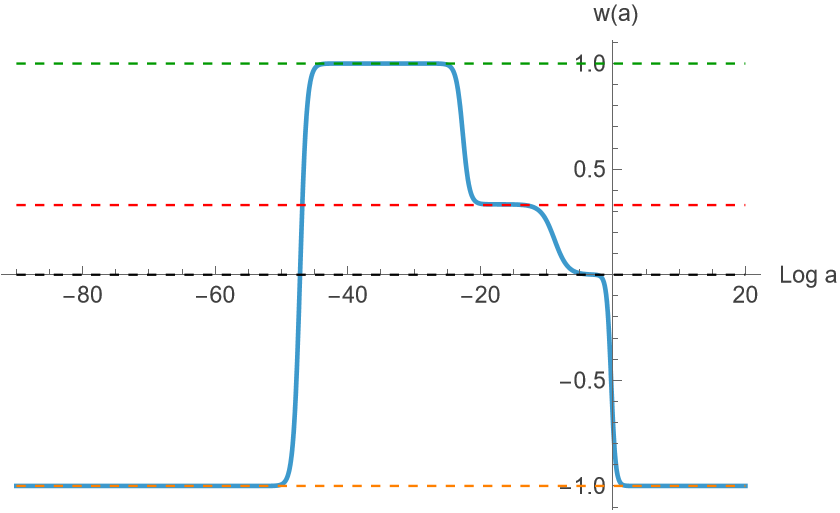}\includegraphics[width=0.5\textwidth]{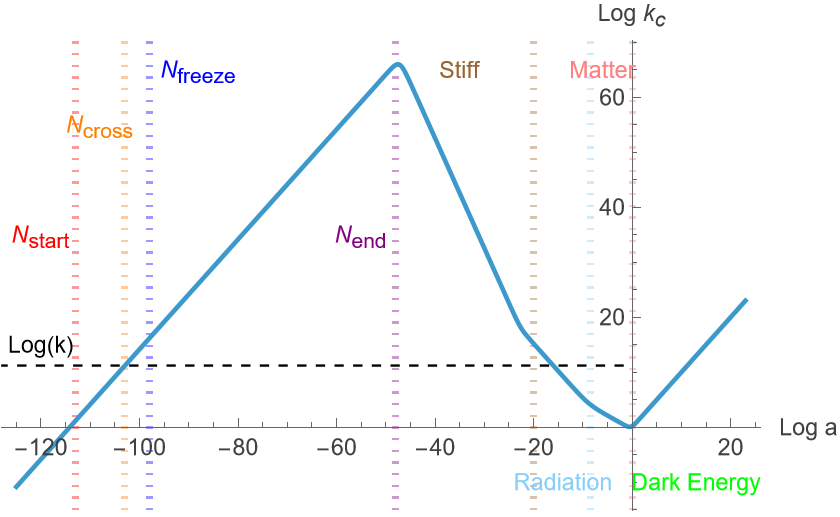}
    \caption{\it  Equation of state { index $w(a)$ and crossing scale function $ k_c(a)=a H(a) $  in the $\Lambda$CDM model. The black dashed line in the right picture  indicates the momentum of a generic mode and its intersection points with the curve correspond to the points where the mode escapes and re-enters the  Hubble horizon of observations.}}
    \label{figadec}
  \end{minipage}
\end{figure}

    For specific choices of the abundaces $\Omega_i$  such universe  undergoes five
    different and well separated eras characterised by flat plateau where the  evolution is dominated by an almost perfect fluid obtained by setting all abundances $\Omega_i$ to zero,  except one. The evolution starts
    with an inflation era ($w=-1$) where the scalar potential energy dominates over the kinetic energy, followed by a reheating era ($w=1$) where kinetic energy prevails, and three subsequent eras  
    eras dominated by radiation ($w=1/3$), matter ($w=0$) and cosmological constant / ``dark energy'' ($w=1$) respectively.  
    A phenomenologically viable choice of the species abundances $\Omega_i$  exhibiting five well separated plateau is given by
   \bea
   \Omega_t &=& {10^{-65}} \qquad , \qquad \Omega_s={10^{-24}} \qquad , \qquad  \qquad \Omega_\gamma =4.6350 \times 10^{-5} \ , \nn\\
\Omega_{\text{m}}  &=& 0.3111    \qquad , \qquad    \Omega_\Lambda  =0.6889 \ . 
   \eea
The equation of state parameter $w(a)$  is then plotted in figure \ref{figadec} as a function of the number of e-folds $\log a$.  The acceleration parameter $-q(a)$, defined as
       \bea
 q(a) &=& -{a \ddot{a} \over \dot{a}^2}   =-1- {a H'(a)\over  H(a) } \ , \label{wqh}
 \eea
is found to have a similar qualitative behavior, which interpolates between different constant values that characterize the dominant fluid component in the corresponding regime.
 We notice that $\Omega_t$ is extremely small, so for $a$ large enough, i.e. away from the initial inflation period, we can discard the $\Omega_t$ term, leading to a multi-component fluid universe with Hubble function  
 \be
   H(a) \underset{\Omega_t \to 0}{\approx}  H_0  \left( a^{-6} \Omega _s  + a^{-4} \Omega _{\gamma } + a^{-3} \Omega _m +    \Omega _{\Lambda } \right)^{1\over 2} \ .
     \ee

Note however that both $\Omega_t$ and $\Omega_s$ though extremely smaller than the other abundances  are needed in order to produce inflation and the phase governed by the stiff matter equation of state $w=1$ where `reheating' (Hot Big Bang) should take place.  
\section{Integrable models of Inflation}
In this section we discuss some interesting choices of $H(\phi)$ leading to phenomenologically appealing models of inflation. Each of those will have some resembling features to models which are widely studied in the literature, as well as some differences that we'll try to test at the level of their predictions for the main inflationary observables. The key advantage of the models we analyze here is that they are integrable by construction, as their dynamical evolution is completely specified in terms of  the Hubble function $H(\phi)$. This will allow us to construct analytic classical inflationary backgrounds, and subsequently study the equations that govern the associated cosmological perturbations.
\subsection{ Integrable Starobinsky  model } 
This model is inspired by the original Starobinsky inflationary  model \cite{Starobinsky:1980te}, which made use of higher derivative corrections to the Einstein equations to produce cosmic acceleration. The model can be derived from Einstein gravity coupled to a single inflaton field $\phi$ with a scalar potential of the form \cite{Kofman:1985aw}
\begin{equation}\label{Starobinsky_V}
V(\phi) \,=\, \frac{3}{4}M^2\,\left(1-e^{\sqrt{\frac{2}{3}}\phi}\right)^2 \ ,
\end{equation}
where $M$ is an appropriate mass scale that sets the scale of inflation by parametrizing the height of the inflationary plateau (for large negative $\phi$). The minimum $V=0$ is conveniently located at $\phi=0$. The same effective description associated with the scalar potential in \eqref{Starobinsky_V} is also found to follow from $R^2$ corrections to the Einstein Hilbert action. In this picture, the Starobinsky model is interpreted as an $f(R)$ modified gravity theory in which the $R^2$ correction is $M^2$ suppressed.
 \begin{figure}[t]
\begin{minipage}{0.5\textwidth}
\centering
\includegraphics[width=1\textwidth]{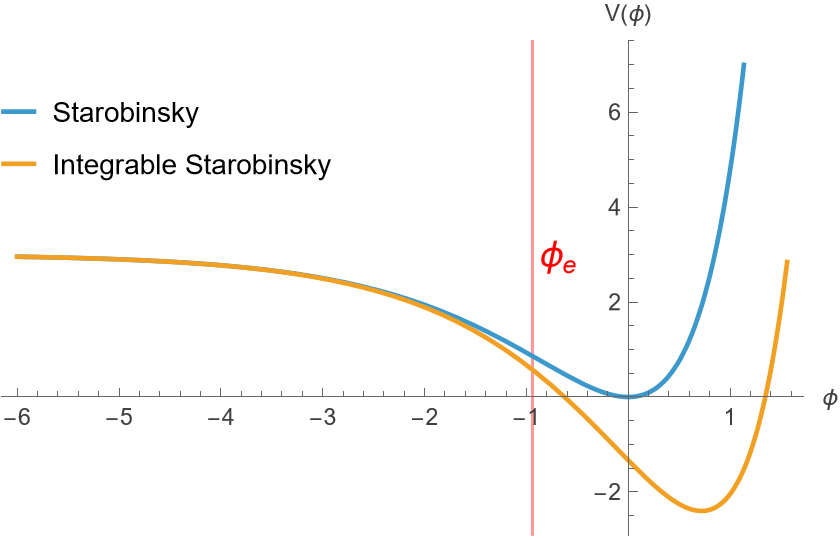}
\end{minipage}
\hfill
\begin{minipage}{0.5\textwidth}
\centering
\includegraphics[width=1\textwidth]{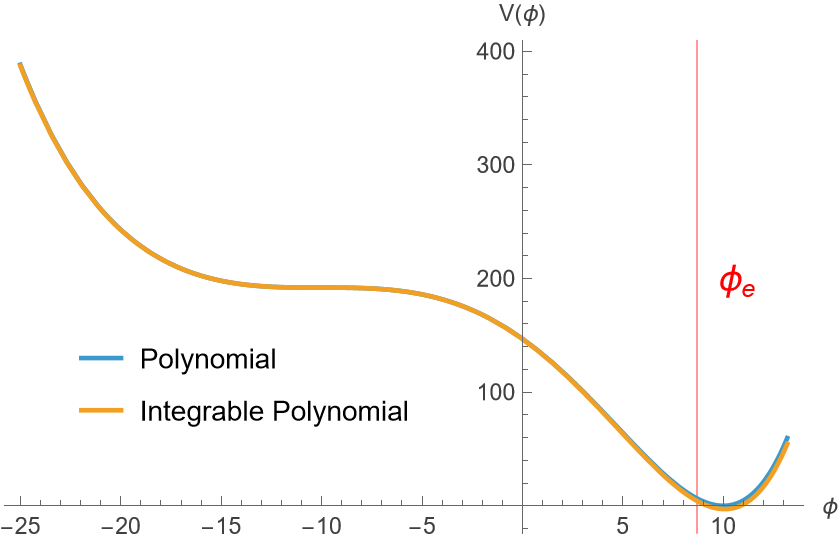}
\end{minipage}
\caption{\it Scalar potentials for the Starobinksy  model  with $\alpha=\sqrt{2/3}$ and the Polynomial model with $\f_0=-1$, $\beta=2$ and $\g=0.1$. }\label{fig:Starobinskypotential}
\end{figure}
Now, a Starobinsky-like model of inflation corresponds to the choice of Hubble function
  \be
  H(\phi)= H_{\rm in} (1- e^{\alpha \phi}) \ ,
  \ee
 leading to the potential
  \be
  V(\phi)=H_{\rm in}  \left[3 \left(1-e^{\alpha  \phi }\right)^2-2 \alpha ^2 e^{2 \alpha  \phi }\right] \ .
  \ee
     In figure \ref{fig:Starobinskypotential} we display the potential for the Starobinsky ispired choice $\alpha=\sqrt{2/3}$, the main difference to the actual Starobinsky potential being the non-zero (in fact negative) value of $V$ at the minimum. Given 
     $H(\phi)$ we can integrate $a(\phi)$, leading to
  \be
 a(\phi)=  e^{  -{1\over 2  } \int  {H(\phi) d\phi\over H_\phi(\phi) } }=  e^ { -\frac{\phi}{2 \alpha  } -\frac{  e^{-\alpha  \phi }   }{2 \alpha ^2 }   }  \ . \label{aphi}
 \ee
  This formula can be then inverted as  
     \be
e^{-\alpha  \phi(a)} =-{ \mathcal{W}_{-1}\left(-a^{2 \alpha^2} \right) }\ ,
 \ee
    where  $ \mathcal{W}_{-1}(x)$ denotes the product-log Lambert function in the $(-1)$-branch. We remind the reader that the Lambert function in the $(-1)$ branch is defined as the smallest root of the two solutions of the equation $ \mathcal{W}(x) e^{ \mathcal{W}(x)}= x$, in the interval $-e^{-1}<x <0$.   
% \footnote{The principal branch admits a series expansion around the origin of the form 
%$ {\cal W}_0(z) = \sum_{n=1}^\infty {(-n)^{n{-}1} \over n!} z^n$.}
The $(-1)$ branch admits an asymptotic expansion around the origin of the form
\be
{\cal W}_{-1}(z) = \log(-z) - \log(-\log(-z))+ ...
\ee 
The fake superpotential as a function of $a$ reads
 \be
 H(a)=H_{\rm in}  \left[ 1+\frac{1}{\mathcal{W}_{-1}\left(-a^{2 \alpha^2} \right)} \right] \ .
 \ee
  \subsection{Polynomial inflation }
Next, we consider a polynomial fake superpotential. We look for an associated potential displaying a sufficiently flat plateau, \emph{e.g.} a point where $V_\phi=V_{\phi\phi}=0$. Using (\ref{vh}), we see that this is garanteed if $H_\phi=H_{\phi\phi}=0$ at some point. 
  The simplest choice allowing for enough freedom to meet the above requirements is a cubic $H$, namely
     \be
   H(\phi) = H_{\rm in} \left[\beta^3-\gamma^3(\phi-{\phi_0})^3 \right] \ ,
   \ee  
    which has a plateau around $\phi={\phi_0}$. The associated potential is given by
    \be
  V(\phi)=3 H_{\rm in}^2 \left[\left(\beta ^3-\gamma^3(\phi-{\phi_0})^3\right)^2-6 \gamma^6(\phi-{\phi_0})^4\right] \ .
    \ee
    We display a plot of this potential in figure \ref{fig:Starobinskypotential} (Right).     Integrating the  equation for $a$, we find
    \be
    a(\phi)=e^{-\frac{\beta ^3}{6\g^3 (\phi-{\phi_0})}-\frac{1}{12} (\phi-{\phi_0})^2}\ ,
    \ee
     and inverting this, one finds
     \be
     \phi = \phi_0+ \frac{1}{\g}\left( \sqrt{\beta ^6 +64 \g^6\log ^3(a)}-\beta ^3 \right)^{1\over 3} -4\g \log (a)  \left( \beta ^6+\sqrt{64\g^6\log ^3(a)}-\beta ^3 \right)^{-{1\over 3} } \ , 
     \ee
  which, in turn, leads to a fake superpotential of the form
   \be
 H(a)=\beta^3 -\left[   \left( \sqrt{64 \g^6 \log ^3(a)+\beta ^6}-\beta ^3 \right)^{1\over 3} -4 \g^2\log (a)  \left( \sqrt{64\g^6 \log ^3(a)+\beta ^6}-\beta ^3 \right)^{-{1\over 3} } \right]^3 \ .
   \ee

 \subsection{ $\alpha$-attractor like models}
The original $\alpha$-attractor models \cite{Kallosh:2013hoa,Ferrara:2013rsa,Kallosh:2013yoa,Galante:2014ifa,Kallosh:2015zsa,Kallosh:2019eeu,Kallosh:2019hzo} arise from polynomial potentials with non-canonical kinetic terms  
 characterised by a hyperbolic geometry of the moduli space. In the canonical variable $\phi$, the potential is  polynomial in the trascendental function $e^{-\alpha \phi}$,  featuring a plateau suitable for inflation.

\begin{figure}[t]
\begin{minipage}{0.5\textwidth}
\centering
 \includegraphics[width=1\textwidth]{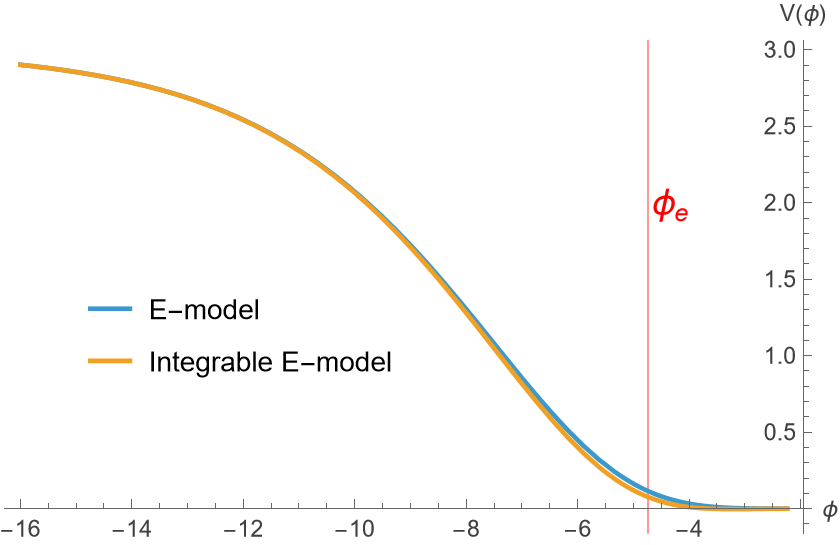}
\end{minipage}
\hfill
\begin{minipage}{0.5\textwidth}
\centering
\includegraphics[width=1\textwidth]{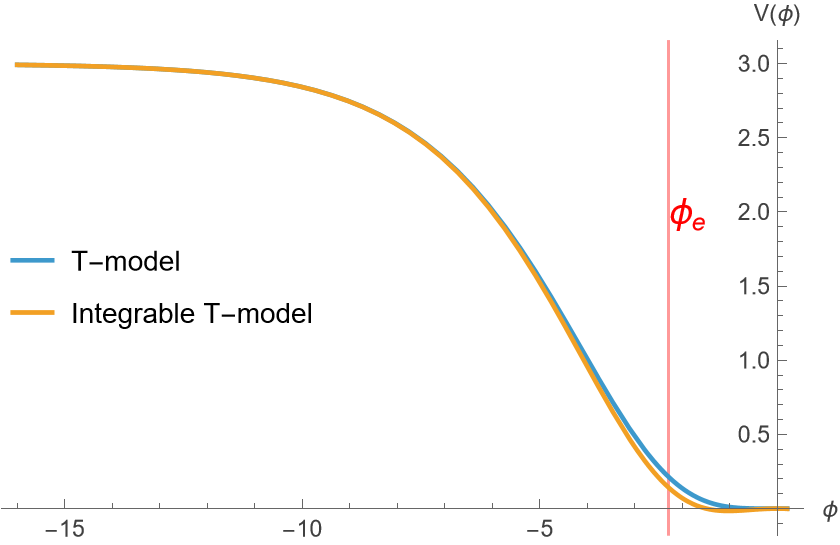}
\end{minipage}
\caption{\it  Scalar potential for alpha attractor models:  E-model with $\alpha=0.4$, $n=10$ and  T-model for $\a=-0.25$ and $n=2$.  }\label{fig:alphaattractors}
\end{figure}

 \subsection*{E-models} 
 
%Let us now move to considering more general inflationary models naturally arising from non-canonical kinetic terms for the inflaton field. In particular, the so-called 
%E- and  $\alpha$-attractor models \cite{Kallosh:2013hoa,Ferrara:2013rsa,Kallosh:2013yoa,Galante:2014ifa,Kallosh:2015zsa,Kallosh:2019eeu,Kallosh:2019hzo} exploit the hyperbolic geometry of moduli space. The non-canonical field is chosen to have a power-law potential, which is transformed into a trascendental function for the canonically normalized inflaton, featuring a plateau suitable for inflation.
The E-models are generalizations of the Starobinsky-like model specified by the following Hubble function
  \be
  H(\phi)= H_{\rm in} (1- e^{\alpha \phi})^{2n}\ ,
  \ee
with $n$ real and positive. The potential reads
  \be
  V(\phi)=H_{\rm in}^2\left(1- e^{\alpha  \phi }\right)^{4 n-2} \left[3 \left(1- e^{\alpha  \phi }\right)^{2}-8 \alpha ^2 n^2 e^{2 \alpha  \phi } \right]\ .
  \ee
  A sketch of the {scalar} potential in this class may be found in  the left figure \ref{fig:alphaattractors}
     Integrating the field equation for $a(\phi)$, one finds
  \be
 a(\phi)=     e^ { -\frac{\phi}{4n \alpha  } -\frac{  e^{-\alpha  \phi }   }{4 n\alpha^2}   }   \ .\label{aphi}
 \ee
As before, this formula can be inverted as  
     \be
e^{-\alpha  \phi(a)} =- \mathcal{W}_{-1}\left(-a^{2 n  \alpha^2} \right) \ ,
 \ee
   leading to
 \be
 H(a)=H_{\rm in}\left[ 1+\frac{1}{\mathcal{W}_{-1}\left(-a^{2 n\alpha^2} \right)} \right]^{2n} \ .
 \ee
\subsection*{T-models}
 {T-models are $\alpha$}-attractor models characterized by a hyperbolic tangent potentials $V\sim \tanh^{2n}\phi$, which enjoy two symmetric plateaux at both infinities.
 {Its integrable version is specified by the Hubble function}
   \be
  H(\phi)= H_{\rm in} \tanh^n(\alpha  \phi ) \ ,\label{hphit}
  \ee
where $n$ is real and positive, leading to the potential
  \be
  V(\phi)=H_{\rm in}^2 [3 \tanh ^{2 n}(\alpha  \phi )-2 \alpha ^2 n^2 \text{sech}^4(\alpha  \phi ) \tanh ^{2 n-2}(\alpha  \phi )]
  \ee
  A sketch of the {scalar} potential in this class may be found in  the right figure \ref{fig:alphaattractors}.
  The profile for $a(\phi)$ reads
  \be
 a(\phi)= e^{-\frac{\cosh (2 \alpha  \phi )}{8 \alpha ^2 n}} \ ,\label{aphi}
 \ee
   or equivalently
   \be
\cosh (2 \alpha  \phi ) =-8 \alpha ^2 n  \log(a) \ . \label{aphi}
 \ee 
  Plugging this expression for $\phi$ into (\ref{hphit}), one finds
 \be
 H(a)=H_{\rm in} \left(\frac{1+8\alpha^2n\log^2{(a)}}{1-8\alpha^2n\log^2{(a)}}\right)^{n/2}\ .
 \ee    
  
  \subsection{Trigonometric models }

 \begin{figure}[t]
\begin{minipage}{0.5\textwidth}
\centering
\includegraphics[width=1\textwidth]{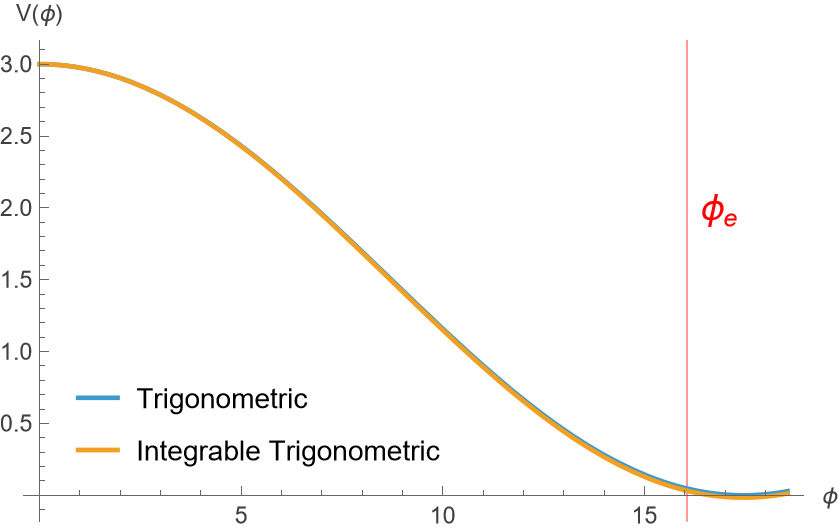}
\end{minipage}
\hfill
\begin{minipage}{0.5\textwidth}
\centering
\includegraphics[width=1\textwidth]{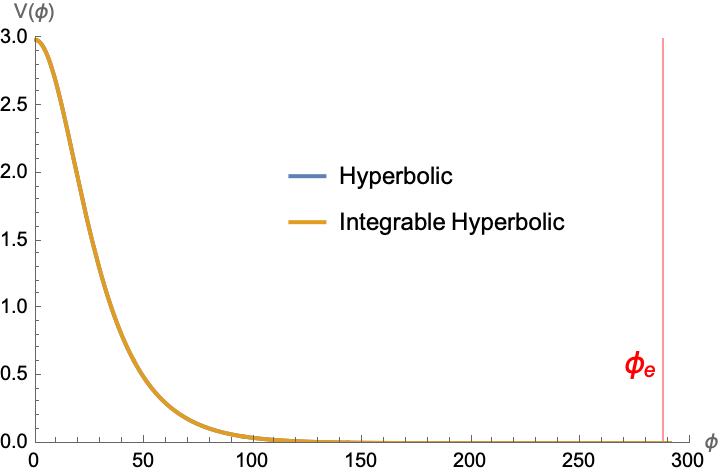}
\end{minipage}
\caption{\it Scalar potentials for the  Trigonometric model with $\alpha=1/10$ and the Hyperbolic model with $\alpha=0.05$, $n=0.5$,.   }\label{fig_hypertrig}
\end{figure}

  Another class of models that we consider is based on trigonometric functions. These models may be seen as generalizations of the well-known natural inflation model of \cite{Freese:1990rb}, which is specified by a $\cos^2\phi$ potential. In our case we start by choosing
   \be
   H(\phi)= \cos ^n(\alpha  \phi ) \ ,
  \ee
 with $n$ real and positive, leading to the potential
  \be
  V(\phi)= 3  \cos ^{2 n}(\alpha  \phi )-2 \alpha ^2  n^2 \tan ^2(\alpha  \phi ) \cos ^{2 n}(\alpha  \phi ) \ .
   \ee
A sketch of the {scalar} potential in this class may be found in  the left  figure \ref{fig_hypertrig}.
    The equation for $a$ is integrated as
    \be
    a= \sin^{\frac{1}{2n\alpha^2}} (\alpha  \phi ) \ ,
    \ee
    leading to
\be
\sin \left(\alpha \phi(a)\right)=a^{2 \alpha ^2 n} \ .
\ee
    Inverting the above relation, one finds the following $H$ profile
    \be
    H(a)=  \left(1-a^{4 \alpha ^2 n}\right)^{n/2}\ .
    \ee

  \subsection{Hyperbolic models }
To conclude our list of integrable inflationary models analyzed in this paper,  let us consider  a hyperbolic fake superpotential \footnote{The viable models defined through this choice have the problem of infinite inflation. The problem can be solved summing a little constant to this H, but this does not allow to analytically invert the relation between $a$ and $\f$ anymore. Since the constant is very little in our work range of values we can analytically invert considering it 0.}
   \be
   H(\phi)= {1 \over  \cosh ^{n}(\alpha  \phi )} \ ,\label{hphicosh}
  \ee
  with $n$ real and positive,  leading to the following scalar potential
  \be
  V(\phi)= 3  \cosh ^{-2 n}(\alpha  \phi )-2 \alpha ^2  n^2 \tanh ^2(\alpha  \phi ) \cosh ^{-2 n}(\alpha  \phi )\ .
   \ee
   A sketch of the {scalar} potential in this class may be found in  the right figure \ref{fig_hypertrig}.
    The equation of motion for $a$ is integrated as
    \be
    a=\sinh^{\frac{1}{2n\alpha^2}} (\alpha  \phi ) \ ,
    \ee
leading to
\be
\sinh\left(\alpha \phi(a)\right)= a^{2 \alpha ^2 n} \ .
\ee
Plugging this into (\ref{hphicosh}), one finds 
\be
H(a)=  \left(1+a^{4 \alpha ^2 n}\right)^{-n/2}\ .
\ee
 A particularly interesting case of this class of models corresponds to the choice $n=3$, $\alpha=1/\sqrt{6}$ leading to
\be
H(a)=  \left(1+a^{2}\right)^{-3/2} \ .
\ee
  This particular model interpolates between a de Sitter vacuum $H(a) \underset{a\to 0}{\sim} 1$ and a universe dominated by stiff matter $H(a) \underset{a\to \infty}{\sim} a^{-3}$. Note that the corresponding energy density $\rho(a)$ scales as $H^2$ and hence behaves as $a^0$ and $a^{-6}$, respectively. It may be worth noticing here that a universe dominated by stiff matter has equation of state $w=1$ (following from equation \eqref{n2w}), which is typically realized by a scalar field in its phase of kinetic dominance. In this light, this toy model could provide a simple realization of a transient between the inflationary and a reheating phase of the early universe, where the inflaton starts releasing its kinetic energy and decays into other particles.  
   
\section{Cosmological perturbations}

The equations of scalar and tensor perturbations around a general FLRW metric have been derived by Muhkanov and Sasaki in \cite{Sasaki:1986hm,Mukhanov:1988jd}. {In this section we review the derivation of these equations and the computations of the Power spectrum and the spectral indices, for a general background specified by the Hubble function $H(a)$ in the time coordinate $a$. }

%\subsection{The slow roll parameters}
%
%      The equations of cosmological perturbations can be conveniently written in terms of the slow roll parameters\footnote{In the cosmological time variable
%    \be
%\epsilon_H=    -{\dot{H}\over H^2}   
%~ , \qquad \eta_H  =-{\ddot{H} \over 2 \dot{H} H}  ~, \qquad 
%  \xi_H^2=  -  {1\over H^2} {d\over dt}\left( H \eta \right)      \ee
%   } 
% \be
%\epsilon_H ={2   H_\phi^2 \over H^2}   
%~ , \qquad \eta_H ={2   H_{\phi\phi}\over H}   ~ , \qquad
%  \xi_H^2=  {4   H_{\phi\phi\phi} H_\phi\over H^2}    \label{slowp}\\
%\ee
%Alternatively,  using conformal time $\eta$, or the universe size $a$ as time variables, the slow roll parameters can be written as
%   \bea
%\eta: \qquad  &&  \epsilon_H  =  -{H_\eta\over a H^2}  ~, \qquad 
%\eta_H= \frac{1}{2}-{H_{\eta\eta} \over 2 a H H_\eta}  ~ , \qquad\qquad  \xi_H^2
%=   -{1\over a H^2} {d\over d\eta}( \eta_H H) \nn\\
%a:  \qquad  && \epsilon_H  =  -{ a H_a\over H}  ~, \qquad 
%\eta_H= -\frac{1}{2}-{a H_a \over 2 H } -{a H_{aa} \over 2 H_{a} } ~ , \qquad\qquad  \xi_H^2
%=   -{a\over H} {d\over d a }( \eta_H H) 
%   \eea 
% We remark that although we use slow roll parameters, we will always write exact equations for the perturbations\footnote{We remind that for universes satisfying  the slow roll conditions: 
% $\dot{\phi}^2\ll V$,  $\ddot{\phi} \ll  V_\phi$, the  slow roll parameter can be written as derivatives of the potential   
%\bea
%\epsilon_V\equiv {  V^2_\phi  \over 2 V^2  }   \approx \epsilon_H  ~ , \qquad\qquad \eta_V \equiv {  V_{\phi\phi}\over V }  \approx \eta_H +\epsilon_H
%\eea
%  }
%   

      \subsection{Scalar perturbations} 
   
 Scalar perturbations can be written as  
\bea
ds_4^2  &=&  a(\eta)^2 \left[ -\left(1+2 \delta\Phi(\eta,\bold{x}) \right) d\eta^2 + \left(1-2 \delta\Phi(\eta,\bold{x}) \right) d \bf{x}^2 \right], \nn\\
\varphi (\eta,{\bf x}) &=&  \phi(\eta) +\delta\varphi(\eta,\bold{x}),
\eea
 with
 \be
\phi' =  \frac{       \sqrt{ 4 (a')^2- 2 a  a''}}{a}  ~ ,% \qquad\qquad   V = \frac{   \left(a a''+(a')^2\right)}{a^4}  
\label{solback}
 \ee
  {the background solution}  
  and  $\delta\Phi$, $\delta\varphi$ characterizing the  perturbations. 
 Writing
 \be
 \delta\Phi(\eta,x)=\int {d^3 k\over (2\pi)^3}\,  e^{+{\rm i} {\bf k}{\bf x} } \Phi_k(\eta) ~,  \qquad\qquad  \delta\varphi(\eta,x)=\int {d^3 k\over (2\pi)^3}\,  e^{+{\rm i} {\bf k}{\bf x} } \varphi_k(\eta)\ ,
 \ee
 and expanding Einstein equations to linear order in the perturbations  one finds
     \bea
&&  \varphi_k - \frac{  2   \partial_\eta \left(a \Phi_k\right)}{a \phi'  } =0\ , \label{eqphieta}\\
 &&    \Phi_k ''+\frac{\Phi_k ' \left(8 \left(a'\right)^3-7 a a' a''+a^2 a^{(3)}\right)}{2 a \left(a'\right)^2-a^2 a''}+\Phi_k  \left(k^2+\frac{\left(a'\right)^2
   a''-2 a \left(a''\right)^2+a a' a^{(3)}}{2 a \left(a'\right)^2-a^2 a''}\right)=0\ . \nn
    \eea
    The first equation determines  $\varphi_k $, while the second equation can be written in the Mukhanov-Sasaki form
     \be
    u_k''(\eta)+ \left( k^2- {z''(\eta) \over z(\eta) } \right) u_k(\eta)=0\ ,  \label{equ}
    \ee
    with
     \be
    u_k =   \frac{  2   \partial_\eta \left(a \Phi_k\right)}{ \phi'  } +{a^2 \phi' \Phi_k \over a'}\ ,   \label{uka}
    \ee
    and\footnote{We notice that $z$ is define up to a rescaling by an arbitrary constant.} 
     \be
      z =  {a^2 \phi' \over a'   } =    a\sqrt{ 4 - {2a a''\over (a')^2}}\ . \label{uzs}
    \ee
    {Indeed, substituting (\ref{uka}) and (\ref{uzs}) into (\ref{equ}) and using (\ref{eqphieta}) one can easily check that the equation is verified. }
    % Equation (\ref{equ})  can be easily checked by substituting (\ref{uzs}) into it,  and verify it holds using (\ref{eqphieta}). 
    
%     We notice that (\ref{equ}) takes the form of a Schrodinger equation with energy $k^2$ and potential
%      $z''/z$. The later can be written in terms of the slow rolling parameters as
%    \be
%    {z''\over z}=a^2 H^2 (2+2\epsilon_H -3 \eta_H +2 \epsilon_H^2  -4 \eta_H\,   \epsilon_H  +\eta_H^2+\xi_H^2 )
%    \ee
%       We remark that this formula is exact, with $a$, $H$ and slow roll parameters all varying with $\eta$. 
       
%    We can write (\ref{uzs}) as
%    \be
%    z_S(\eta)= a \sqrt{2\zeta(\eta)}
%    \ee
%    with $N=\log a$ and
%    \be
%     \zeta= 1- {N''\over (N')^2 }
%    \ee 
%    the slow roll function. 
    
       \subsection{Tensor perturbations} 
   
 Tensor perturbations are described by the metric
\bea
ds_4^2  &=&  a(\eta)^2 \left[ - d\eta^2 +  d {\bf x}^2 + \delta h_{ij}(\eta,{\bf x} ) dx^i dx^j  \right]\ , \nn\\
\varphi (\eta,{\bf x}) &=&   { \phi(\eta)  }\ ,  
\eea
 with $\phi(\eta)$ following again from (\ref{solback}) and  {  $\partial^i\delta h_{ij}=0$. We denote by $\delta h=\ft12 \sum_{ij} (\delta h_{ij})^2$ and write}
  \be
 { \delta h (\eta,{\bf x}) = \frac{2}{a(\eta)}  \int {d^3 k\over (2\pi)^3}\,    e^{{\rm i} {\bf k}{\bf x} } \, v_k(\eta)  }~,   
 \ee
  with $k^j\xi_{ij}=\xi^i{}_i=0$.   At linear order in the perturbations, the Einstein equations reduce to  
    \be
    v_k''(\eta)+ \left( k^2- {a''(\eta) \over a(\eta) } \right) v_k(\eta)=0\ ,  \label{equ2}
    \ee
%    with
%     \bea
%    v_k(\eta) = {h (\eta) \over a(\eta) }.     
%    \eea
%    In terms of the slow roll parameters one finds
%        \be
%     {a'' \over a}=a^2 H^2 (2-\epsilon_H)
%    \ee

 \subsection{The Power spectrum}
 
  In this section, we reformulate the Mukhanov--Sasaki perturbation equations, together with the power spectra and the spectral indices, in terms of the {scale factor} time variable. The use of this time variable is motivated by the fact that, for the inflationary models considered in this work, the dependence $a(\phi)$ can be integrated analytically and explicitly inverted, yielding a closed-form expression for $H(a)$. 
%   and easier analytical approximations. In addition to this, the $a$ variable will reveal itself convenient for the numerical integration. In some cases it also allows a connection with quantum Seiberg-Witten curves for ${\cal N}=2$ SYM theories with gauge group $SU(2)$ and hypermultiplets in the fundamental \cite{MBGDJFMgmeetc}. \\
   This parametrization makes it possible to trace the evolution of cosmological perturbations throughout the entire inflationary epoch without invoking the slow-roll approximation. 
   %Moreover, as we will see in the next section, these equations can be integrated via numerical and semi-analytic methods.

{ At linear order in the perturbations, using the time variable $a$, we write }
  \bea
ds_4^2  &=& - { da^2\over a^2 H^2} \left[1+2 \delta{\Psi}(a,\bold{x}) \right]  + a^2 \left[1-2 \delta{\Psi}(a,\bold{x}) \right]  {\bf dx}^2+  a^2 {\delta h_{ij} } (a,{\bf x} )   dx^i dx^j\ ,   \nn\\
\varphi (a,{\bf x}) &=&  \phi(a) +\delta\varphi(a,\bold{x})\ .
\eea  
   { with $\partial^i\delta h_{ij}=0$. As before, we denote by $\delta h=\ft12 \sum_{ij} (\delta h_{ij})^2$ and introduce the gauge invariant Fourier components}
    \bea
   u_k(a) &=&   \int d^3 k \,  e^{-{\rm i} {\bf k}{\bf x} } \, \left[ \frac{  2   \partial_a \left(a \delta {\Psi}\right)}{ \partial_a \phi   } + a^2  \partial_a \phi  \, \delta {\Psi}    \right]\ ,  \nn\\
 2 v_k(a) & =&  a  \int d^3 k \,  e^{-{\rm i} {\bf k}{\bf x} } \,  \delta h(a,{\bf x} ) \ .
 \eea
 
     The equations for scalar and tensor perturbations can be found from (\ref{equ}) and (\ref{equ2})
   translating $\eta$-derivatives into $a$-derivatives and can be written in the unified form
    \be
  \mathfrak{u}_{k}''(a)+\left(  k^2 \, J(a)^2  -{ {\cal Z}''(a)\over{\cal Z}(a) } \right)    \mathfrak{u}_{k}(a)=0 \ ,  
 \label{equk2bis}
\ee
   with   
  \be
 \mathfrak{u}_{k}(a)=\left\{ 
\begin{array}{c}
  {u_{k}(a)\over \sqrt{J(a)}}   \\
 {2v_{k}(a)\over \sqrt{J(a)}} \\
\end{array}
\right. ,  \qquad\qquad  {\cal Z}(a)=\left\{ 
\begin{array}{c}
  \ft{z(a)}{ \sqrt{J(a)}}  \\
 \ft{ a}{ \sqrt{J(a)}}    \\
\end{array}
\right. =\left\{ 
\begin{array}{ccc}
  a^2  \sqrt{ -2 a H'(a) } &~~~~~& {\rm Scalar} \\
 a^2 \sqrt{   H(a) } &~~~~~& {\rm Tensor} \\
\end{array}
\right. ,    \label{zja}
  \ee 
   and
  \be
  J(a)=\eta'(a)={1\over a^2 H(a)}\ .  
   \ee

 We look for solutions satisfying Bunch-Davies boundary conditions at the earliest times $k\eta \to -\infty$  
\be
{ \mathfrak{u}_k(\eta) } \underset{k\eta \to -\infty}{\to} {1\over \sqrt{2k} } e^{-{\rm i} k \eta }\ , 
\ee
or equivalently in the {scale factor} time variable, 
\be
\mathfrak{u}_{k}(a)\underset{k\gg a {H_{\rm in}} }{\to} {a \over \sqrt{2 k } } e^{ {{\rm i} k\over a H_{\rm in} }  }\ ,   \label{bdbc}
\ee
with $H_{\rm in}=H(0)$. 
 In the opposite limit, one finds 
  \be
  { \mathfrak{u}_{k}(a)  \over{\cal Z}(a)  } \underset{ k\ll a {H_{\rm in}} }{\sim} \mathfrak{c}(k)\ ,   
 \ee
 with  $\mathfrak{c}(k)  $ some constant independent of $a$.   The  power spectra  are defined  as\footnote{
  The power spectrum of a field  $\Psi(\eta,{\bf x}) = \int {d^3 k\over (2\pi)^3}\,  e^{{\rm i} {\bf k}{\bf x} } \, \Psi_k(\eta)$  is defined in terms of the quantum two-points correlator  
 \be
   \int dx \langle  \Psi(\eta,{\bf x})^2 \rangle= \int {d^3k \over (2\pi)^3} \left|  \Psi_k(\eta)  \right|^2 = \int  {dk\over k} P(k,\eta) . 
 \ee
 }
\be
\label{eq:Power spectra definition}
{P_{S}(k,a) = {k^3\over 2 \pi^2} \left|  { u_k(a)  \over z(a)  }  \right|^2,   \qquad   P_{T}(k,a) =2\times  {k^3\over 2 \pi^2} \left|  { 2v_k(a)  \over a }  \right|^2 } \ ,
\ee
 { where the extra 2 accounts for the two degrees of freedom of tensor perturbations.}
% For each $k$, we denote by ${\cal P}(k)$ the value $P(k,\eta_k)$  at the time $\eta_k$  when the mode $k$ crossed the horizon, i.e. 
%\be
%{\cal P}_{S,T}(k) = P_{S,T}(k,\eta_k) ,  \qquad {\rm with} \qquad k =a(\eta_k) H(\eta_k) 
%\ee 
%  We introduce the amplitudes 
% \be
% A_S(k)= \ft{2}{5} {\cal P}_S(k)^{1\over 2}  , \qquad \qquad    A_T(k)= \ft{1}{10} {\cal P}_T(k)^{1\over 2}
% \ee
 In terms of these, the spectral indices are given by 
\be
 r(k)  = {P_T\over  P_S} ~, \qquad\qquad  n_S(k) = 1+{d \log P_S\over d\log k } ~, \qquad\qquad  n_T(k)={d \log P_T\over d\log k }\ ,  \nn\\
\ee
 where all quantities are computed at late times $k  \ll a {H_{\rm in}} $  when the power spectrum is completely frozen to a constant value {that depends on}  $k$.

     \subsection{One component universe}
    
    Let us consider first a one component universe with prepotential
    \be
    H(a) =  H_{\rm in} a^{-{n \over 2} }\ . 
   \ee
  Plugging in this into (\ref{zja}) one finds
   \be
   {\cal Z}(a)=\left\{ 
\begin{array}{ccc}
  \sqrt{ n} \, a^{2 -{n \over 4}   }   &~~~~~& {\rm Scalar} \\
 a^{2-{n\over 4} }  &~~~~~& {\rm Tensor} \\
\end{array}
\right. ,   
  \ee 
    leading to
    \be
     {{\cal Z}''(a)\over{\cal Z}(a)}= {(8-n)(4-n)\over 16 a^2}\ .
    \ee
   The scalar and tensor perturbation equations take  the {same}  form
   \be
  \mathfrak{u}_{k}''(a)+\left( {k^2\over H_{\rm in}^2 a^{4-n} }   -{ \nu^2-\ft14 \over  a^2 }\right) \mathfrak{u}_{k}(a)=0\ ,   \nn\\
 \label{equk2}
\ee
%   
%   The conformal time coordinate is given by
%    \be
%  \eta(a) =   \int {da\over a^2 H(a)}=  \ft{2}{n-2}a^{\frac{n-2}{2}} 
%  \ee
%leading to
%    \bea
%   a(\eta)=\left(\ft{n-2}{2} \eta \right)^{2\over n-2} 
%  \eea
%   Plugging this into (\ref{uzs}) one finds
%    \be
%   {z''(\eta) \over z(\eta) }= {a''(\eta) \over a(\eta) }=\frac{2 (4-n)}{\eta ^2 (n-2)^2}
%   \ee 
%     The scalar and tensor perturbations  are therefore governed by the same Bessel equation
%    \be
%   U''(\eta )+ U(\eta ) \left(k^2-\frac{ \nu^2-\ft14}{ \eta^2  }\right)=0 \label{onec}
%    \ee
    with    
     \be
     \nu= {n-6\over 2(n-2) }\ .
      \ee
    The general solution reads  
%   \be
%    U(\eta)=  c_+ \sqrt{\eta} \, H^{1}_{\nu}( k \eta ) +c_- \sqrt{\eta} \, H^{1}_{\nu}( -k \eta )
%    \ee
%   with $H^1_\nu$ the Hankel function.  Alternatively, using $a$ as a time variable, the perturbation equations (\ref{equk2})  reduce to
%    
%  with $U_c=(u_c,v_c)$ and general solution
  \be
  \mathfrak{u}_{k}(a)= \sqrt{a} \left[ c_+  H_{\nu}^{(1)}\left(\ft{2\, k\, a^{\frac{n}{2}-1}}{(2-n)H_{\rm in} }  \right) +c_-   H_{\nu}^{(2)}\left( \ft{2\, k\,a^{\frac{n}{2}-1} }{(2-n)H_{\rm in} } \right) \right]\ . 
  \ee
  where $H_{\nu}^{(1,2)}$ are Hankel functions of the first and second kind.
It may be worth remarking at this point that generalizations to two fluid components are relevant in the study of cosmological transients within the Hot Big Bang model. In particular, the cosmological perturbations in the transient between radiation and matter domination are described by a Heun equation and may be studied by applying the quantum Seiberg-Witten (qSW) / Cosmology dictionary introduced in \cite{Bianchi:2024mlq}.
   
  \subsection*{De Sitter inflation}
   
   Perturbations around the De Sitter vacuum are described by  (\ref{equk2bis}) with $n=0$,  leading to 
     \be
  \mathfrak{u}_{k}''(a)+\left( {k^2\over H_{\rm in}^2 a^{4} }   -{  2 \over  a^2 }\right) \mathfrak{u}_{k}(a)=0\ ,   \nn\\
 \label{eqdesitter}
\ee
    with general solution
  \be
    \label{eq:De Sitter solution}
    \mathfrak{u}_{k}(a)  =\sqrt{a} \left[ c_+  H_{3\over 2}^{(1)}\left(\ft{k}{a  H_{\rm in} }   \right) +c_-   H_{3\over 2}^{(2)}\left(  \ft{k}{a  H_{\rm in}}  \right) \right].
    \ee 
     Imposing Bunch-Davies boundary conditions (\ref{bdbc}) at $a=0$ one finds
    \be
    \label{eq:De Sitter solution}
    \mathfrak{u}_{k, {\rm in} }(a)=\frac{  \sqrt{ \pi a}}{ 2 } H_{3\over2}^{(1)}\left(\ft{k}{a  H_{\rm in} } \right) =\frac{a(k+ia)}{\sqrt{2 k^3}}e^{\frac{ i k}{a  H_{\rm in} }}.
    \ee

\subsection{The slow-roll approximation}

    We conclude this section by reviewing the standard computation of the power spectra and spectral indices, based on the slow-roll approximation   $|\dot{\f}|^2 \lll |V|$.  The results will be later used for comparisons against those obtained by numerical and semi-analytic integrations of the perturbation equations.  
       
     We start by introducing the basic slow-roll functions in the cosmological time variable
 \be
\epsilon_H(t)=    -{\dot{H}\over H^2}   
~ , \qquad \eta_H(t)  =-{\ddot{H} \over 2 \dot{H} H}  ~, \qquad 
  \xi_H(t)^2=  -  {1\over H^2} {d\over dt}\left( H \eta_H \right).     
 \ee
   Using (\ref{dtime}) we can rewrite these formulae in different time variables\footnote{ We notice that, at leading order in the slow-roll approximation $\dot{\phi}^2\ll |V|$, $|\ddot{\phi}| \ll  |V_\phi|$, 
the  slow-roll functions are nearly constants and their form defined with respect to the potential $V(\phi)$ can be written linearly in terms of expressions defined through the Hubble function
\bea
\epsilon_V\equiv \frac{  V^2_\phi}{2 V^2  }   \approx \epsilon_H  ~ , \qquad\qquad \eta_V \equiv \frac{  V_{\phi\phi}}{V}  \approx \eta_H +\epsilon_H
\eea  
 }:
  \bea
\phi: \qquad  &&  \epsilon_H   ={2   H_\phi^2 \over H^2}  ~, \qquad 
\eta_H = {2   H_{\phi \phi}\over H }   ~ , \qquad
  \xi_H^2=  {4   H_{\phi\phi\phi} H_\phi\over H^2};  \\
\eta: \qquad  &&  \epsilon_H  =  -{H_\eta\over a H^2}  ~, \qquad 
\eta_H= \frac{1}{2}-{H_{\eta\eta} \over 2 a H H_\eta}  ~ , \qquad\qquad  \xi_H^2
=   -{1\over a H^2} {d\over d\eta}( \eta_H H); \nn\\
a:  \qquad  && \epsilon_H  =  -{ a H_a\over H}  ~, \qquad 
\eta_H= -\frac{1}{2}-{a H_a \over 2 H } -{a H_{aa} \over 2 H_{a} } ~ , \qquad\qquad  \xi_H^2
=   -{a\over H} {d\over d a }( \eta_H H); \nn
   \eea 
% We remark that although we use slow roll parameters, we will always write exact equations for the perturbations\footnote{We remind that for universes satisfying  the slow roll conditions: 
% $\dot{\phi}^2\ll V$,  $\ddot{\phi} \ll  V_\phi$, the  slow roll parameter can be written as derivatives of the potential   
%\bea
%\epsilon_V\equiv {  V^2_\phi  \over 2 V^2  }   \approx \epsilon_H  ~ , \qquad\qquad \eta_V \equiv {  V_{\phi\phi}\over V }  \approx \eta_H +\epsilon_H
%\eea
%  }
%   

For instance, using conformal time as time variable, the scalar and tensor perturbation equations can be written exactly in terms of the slow-roll parameters as
 \be
  \mathfrak{u}_{k}''(\eta)+\left(  k^2   -{{ \cal Z}''(\eta)\over{\cal Z}(\eta) } \right)    \mathfrak{u}_{k}(\eta)=0  \ , 
 \label{equk3}
\ee
     with
  \be
 \mathfrak{u}_{k}(\eta)=\left\{ 
\begin{array}{c}
  u_{k}(\eta)   \\
 v_{k}(\eta) \\
\end{array}
\right. ,  \qquad   {{ \cal Z}''(\eta)\over{\cal Z}(\eta) } =\left\{ 
\begin{array}{l}
  {z''\over z} = a^2 H^2 (2{+}2\epsilon_H {-}3 \eta_H +2 \epsilon_H ^2 { -}4 \eta_H  \epsilon_H  {+}\eta_H^2+\xi_H^2 )   \\
 {a''\over a} = a^2 H^2 (2{-}\epsilon_H)   \\
\end{array}
\right.     \label{zja2}
  \ee 
%The slow roll approximation can be thought of as an approximation around the only known exact solution of the \textit{Power-Law inflation}\footnote{To be correct there is another exact case introduced by Easther \cite{Easther}, but while being theoretically interesting it has been excluded by observations}.  In this exact case the slow roll functions  $\epsilon_H$, $\eta_H$, $\xi_H$ are constant and equal to $\frac{1}{p}$, where $p$ is a constant, and one can use  
{In the slow-roll approximation,  $\epsilon_H$, $\eta_H$, $\xi_H$ are nearly constant, and one can write}
\be
\label{eq:eta in terms of a in exact solution}
\eta=\int\frac{da
}{a^2H(a)}=-\frac{1}{aH(a)}+\int\frac{\epsilon_H da}{a^2H(a)} {=-\frac{1}{aH(a)}+ \epsilon_H \, \eta } \ ,
\ee
{leading to}
\be
{ a H(a) =-\frac{1}{ \eta}\frac{1}{1-\epsilon_H}\ . }
\ee
 Plugging this into (\ref{zja2}) one finds that the perturbation equations reduce to Bessel equations with 
 \be
 \frac{{\cal Z}''(\eta)}{{\cal Z}(\eta)} =\frac{\nu^2-\frac{1}{4}  }{\eta^2}   \ .
   \ee 
%where at first order $\mathfrak{n}=\n=\m=\frac{3}{2}+\frac{1}{p-1}$. These will not be equal anymore when we expand around power law inflation for the other models, and the slow-roll parameters will not be exactly constant.
Using (\ref{zja2}) one finds at linear order in the slow-roll parameters
 \be
 \nu =    \left\{ 
\begin{array}{l}
 \ft{3}{2} {+}2 \epsilon_H {-} \eta_H +\ldots      \\
 \ft32 {+}  \epsilon_H +\ldots  \\
\end{array}
\right.  ,    \label{zja}
  \ee 
The solution satisfying Bunch-Davies boundary conditions at the origin ($\eta\rightarrow -\infty$) reads
    \be
 \mathfrak{u}_k(\eta)= { \frac{\sqrt{\pi}}{2}e^{i(\nu+1/2)\frac{\pi}{2}}(-\eta)^{1/2}H_{\nu}^{(1)}(-k\eta) }\ .
 \ee
   {  In the opposite limit $k \eta \to 0$, where power spectra get {\it frozen}  one finds 
 \be
 \mathfrak{u}_k\underset{ {k\eta  \to 0}}{\sim}  { e^{i (\nu-1/2)\pi/2}2^{\nu-3/2}\frac{\Gamma(\nu)}{\Gamma(3/2)}\frac{1}{\sqrt{2k}}(-k\eta)^{-\nu+1/2} }\ ,
\ee
Plugging this into (\ref{eq:Power spectra definition}), one finds
\be
\label{eq:Power spectra definition}
  \frac{d \log P_{S,T} }{d \log k} = 3-2\nu_{S,T} \ , 
\ee
leading to the slow-roll formulas
\be
 n_S = 1-4 \epsilon_H +2 \eta_H +\ldots ~, \qquad\qquad  n_T=-2 \epsilon_H +\ldots
\ee
 On the other hand,  
 \be
  r(k)  = {P_T\over  P_S}\approx {8z^2\over a^2}+\ldots  =  16 \epsilon_H+\ldots\ ,
 \ee
 where dots denotes corrections to the slow-roll approximation.    
 }

\section{Computations of the power spectra and spectral indices }

In this section, we compute the power spectra and spectral indices of some phenomenologically appealing inflationary models.  We focus on integrable inflationary models, where the Einstein equations for $a(\phi)$ can be integrated and inverted leading to an analytical form for $H(a)$. We then integrate the Mukhanov Sasaki perturbation equations in the time variable $a$, using numerical and semi-analytical methods that do not rely on a slow-roll approximation of the potentials. 
  We stress the fact that the use of the time variable $a$, significantly simplifies the numerical integration with respect to the natural variable $\f$, since the singularities of the equations at the origin are Fuchsian  rather than trascendental.

\subsection{Numerical and semi-analytic integration }

\begin{figure}[t]
\begin{minipage}{0.5\textwidth}
\centering
\includegraphics[width=1\textwidth]{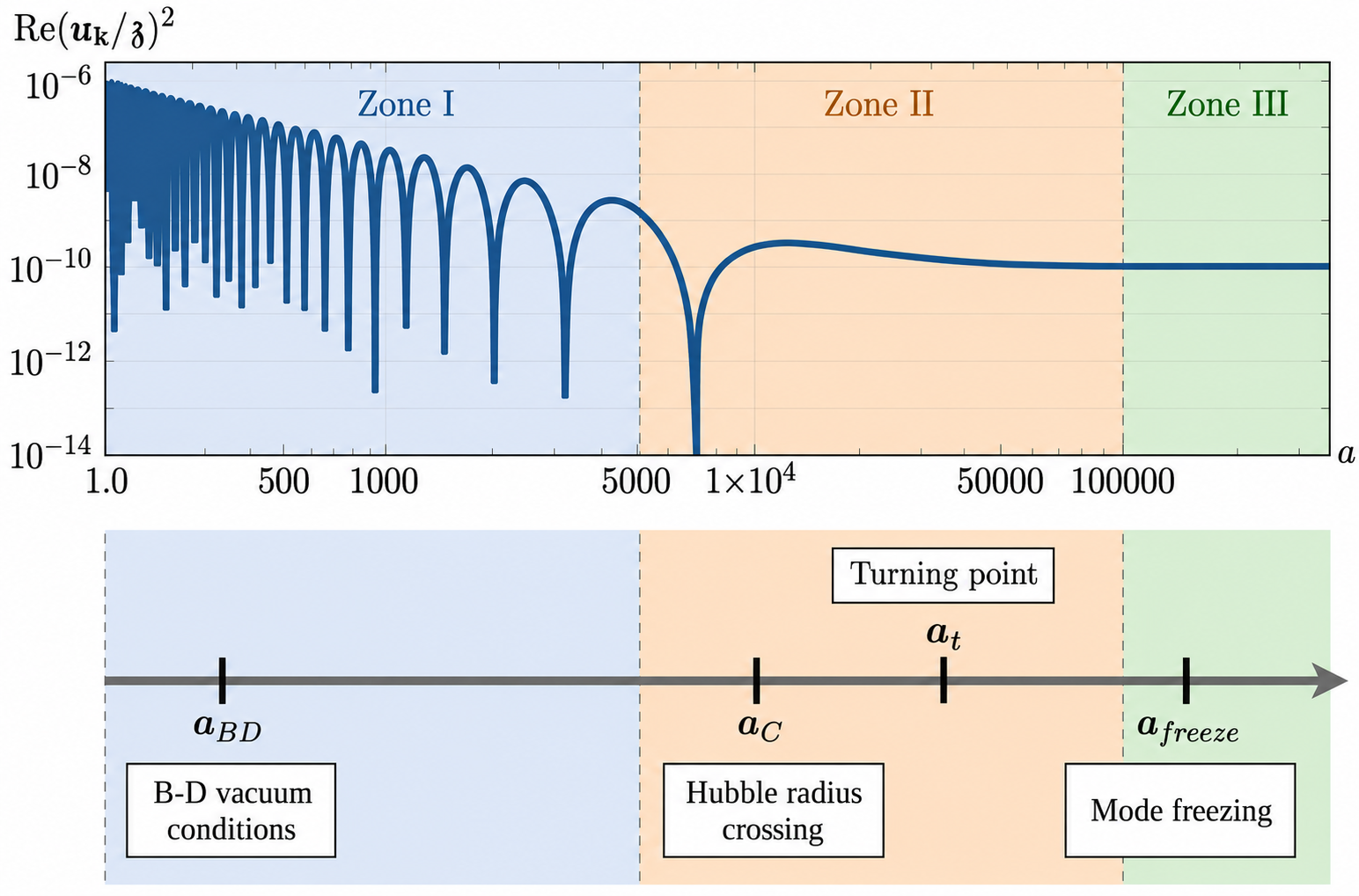}
\end{minipage}
\hfill
\begin{minipage}{0.5\textwidth}
\centering
\includegraphics[width=1\textwidth]{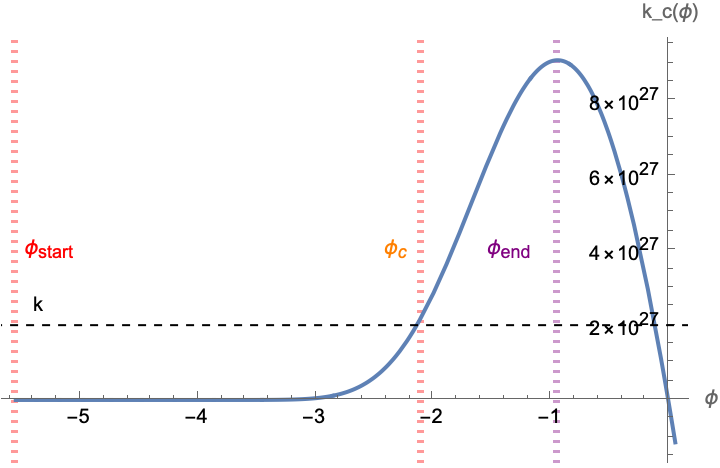}
\end{minipage}
\caption{\it   The cosmological perturbation time line: from the Bunch-Davis vacuum to freezing. In the right figure we display the function $k_c(\phi) = a(\phi) H(\phi)/H_{\rm in}$ that shows the time where a mode of given $k$ escapes from the
Hubble horizon. }\label{fig_bdfreezing}
\end{figure}

Let us first set our conventions and outline the general strategy.  We start from the perturbation equations (\ref{equk2bis}),  {that can be written as}
\be
\mathfrak{u}_{k}''(a)+Q(a) \mathfrak{u}_{k}(a)=0\ ,
\ee
with 
 \be
  Q(a)= {k^2 \over a^4 H(a)^2}    -{ {\cal Z}''(a)\over {\cal Z}(a) } \ ,    \label{zja}
  \ee 
  and
  \be
        {\cal Z}(a) =\left\{ 
\begin{array}{l}
{\cal Z}_S= a^2  \sqrt{ -2 a H'(a) }   \\
{\cal Z}_T= a^2 \sqrt{   H(a) }  \\
\end{array}
\right. ,    \label{zja2}
  \ee 
   We consider pertubations blowing up from a De Sitter universe at the origin of time $a=0$, so we look for solutions of the perturbation equations (\ref{equk2}) around de Sitter geometry, satisfying Bunch-Davis boundary conditions (normalized outgoing waves). More precisely we look for solutions satisfying  
         \be
    \mathfrak{u}_{k }(a) \underset{a\to 0 }{\approx}  \mathfrak{u}_{k, {\rm in} }(a) =\frac{a(k+ia)}{\sqrt{2 k^3}}e^{\frac{ i k}{a  H_{\rm in} }}\ .
    \ee
    We denote by $N(\phi)$ the number of e-foldings of the universe from a given reference point $\f_{\rm start}$ near the origin of inflation. More precisely we write
 \be
 \label{eq:scale with conventions}
 a(\f)=e^{N(\f)} \ ,
 \ee
 with
 \be
 \label{eq:scale with conventions2}
 N(\f)=-\int_{\f_{\rm start}}^{\f}\frac{H\left(\bar{\f}\right)}{2H'\left(\bar{\f}\right)}d\bar{\f}\ .
 \ee
     We can measure the time by $a$, $N$ or $\phi$,  but we find more instructive  to follow evolution tracing  the auxiliary parameter
\be
\kappa(a) ={k\over a\, H(a)}\ .
\ee
 At early times,  $\kappa$ is large, and the $k$-dependent term of $Q(a)$ in (\ref{zja}) dominates producing a rapidily oscillating behavior on the perturbation mode. At very late times $\kappa$ gets small leading to a freezing of the mode. In the middle, the two terms are of the same order. In particular, when $\kappa(a_c)=1$ the perturbation mode crosses the Hubble horizon. After a short time, the power spectrum associated to this mode gets frozen. 
 The complete timeline for a typical mode is displayed in figure \ref{fig_bdfreezing}.

       We remark that $a=0$ is a singular point of the differential equation, so imposing boundary conditions at this point is subtle. Rather than integrating the differential equation from $a=0$, we find it computationally more convenient to start from the
       {\it starting point} $a_{\rm in}$ in the past, several e-foldings from the {\it crossing point} $a_c$, \emph{i.e.} we choose $a_{\rm in}=a_c\,e^{-N_{\rm reg}}$  with $N_{\rm reg}$ chosen large enough. We find that the results are nearly insensitive to the choice of the regulator $N_{\rm reg}$, as long as it is chosen so that  $N_{\rm reg}\geq 5$.  Typically, a few e-foldings after the crossing point, one reaches the  {\it turning point} $a_t$ where $Q(a_t)=0$  and later the  {\it freezing point} $a_{\rm freeze}$, where $\kappa(a_{\rm freeze})=10^{-3}$ and the power spectrum gets frozen to a constant value. 
 
    It is important to remember here that e-foldings are counted from a starting reference point $\phi_{\rm start}$. This starting point is conventionally chosen to be $N_e=65$ from the {\rm inflation ending time} $a_{\rm e} $, \emph{i.e.} the time where acceleration stops $\ddot{a}_{\rm e}=0$, defined as the solution of
  \be
  \epsilon_H(a_{\rm e}) =-{a_{\rm e} H'(a_{\rm e} )\over H(a_{\rm e})} =1 \ .
 \ee

\noindent We follow {three} strategies to solve the differential equations.

\begin{itemize}

\item{ {\bf Numerical Integration}: We numerically integrate  scalar and tensor perturbation equations (\ref{equk2}), in the interval $a\in [a_{\rm in}, a_{\rm freeze}]$ with initial boundary conditions 
\be
\mathfrak{u}_{k}(a_{\rm in})=\mathfrak{u}_{k,{\rm in} }(a_{\rm in}) ~, \qquad\qquad  \mathfrak{u}'_{k}(a_{\rm in})=\mathfrak{u}'_{k,{\rm in} }(a_{\rm in}) \ .
\ee
}
\item{ {\bf Semi-analytic I}:  We approximate the solution by the following  piecewise analytic function
\bea
 \mathfrak{u}_{k}(a) &=& \frac{a(k+ia)}{\sqrt{2 k^3}}e^{\frac{ i k}{a  H_{\rm in} }}\ ,  \qquad\qquad   a \in [0,a^{\rm match}_1] , ~\kappa\gg 1 \nn\\
 \mathfrak{u}_{k}(a) &=&  \sqrt{a} \left[ c_1   H_{\nu }^{(1)}\left(\ft{k}{ a H_{\rm in} }  \right) +c_2   H_{\nu}^{(2)}\left(  \ft{k}{ a H_{\rm in} }\right) \right] ,   \quad  a\in [a^{\rm match}_1,a^{\rm match}_{2}]\ ,  ~\kappa= 1\nn\\
 \mathfrak{u}_{k}(a) &=& c_{ 3 }\, {\cal Z}(a) +c_{ 4 } \,{\cal Z}(a)\int_1^a\frac{db}{{\cal Z}(b)^2}\ ,  \quad a\in  [a^{\rm match}_{2},a_{\rm freeze}], ~\kappa\ll 1
\eea
where $ a^{\rm match}_1 \in  [0,a_c]$,  $ a^{\rm match}_2 \in [a_c,a_{\rm freeze} ]$.
 We analytically solve the equations at early and late times, respectively. Assuming that  $a_c\ll 1$, we approximate the function in the interior, by the local solution in terms of Hankel functions near the crossing point.  
The index of the Hankel functions is given in terms of the slow-roll parameters {
by
  \be
 \nu= \sqrt{ a_c^2 \frac{{\cal Z}''(a_c)}{{\cal Z}(a_c)}+\ft14 }  =  \left\{ 
\begin{array}{l}
  \sqrt{\frac{9}{4}-3\eta_H+\frac{3}{2}\e_H-3\eta_H\e_H+\frac{5}{4}\e_H+\eta_H+\xi_H^2}   \\
 \sqrt{\frac{9}{4}-\frac{3}{2}\e_H+\eta_H\e_H-\frac{3}{4}\e_H^2}  \\
\end{array}
\right.  \ ,
  \ee 
  }evaluated at the crossing point $a=a_c$.    Finally, the coefficients $c_i$ are determined by matching the local solutions and their first derivatives  at the matching points. 
  }

\item{ {\bf Semi-analytic II}:  We approximate the solution by the piecewise defined function
\bea
 \mathfrak{u}_{k}(a) &=& \frac{a(k+ia)}{\sqrt{2 k^3}}e^{\frac{ i k}{a  H_{\rm in} }}\ ,  \qquad\qquad   a \in [0,a^{\rm match}_1] , ~\kappa\gg 1 \nn\\
 \mathfrak{u}_{D}(a) &=& c_1 \,D_{\mu^+}\left[ x(a)  \right]{+} 
\, c_2 D_{\mu^- }\left[  { i} x(a) \right] \quad  a\in [a^{\rm match}_1,a^{\rm match}_{2}]\ ,  ~\kappa < 1 \nn\\
 \mathfrak{u}_{k}(a) &=& c_{ 3 }\, {\cal Z}(a) +c_{ 4 } \,{\cal Z}(a)\int_1^a\frac{db}{{\cal Z}(b)^2},  \quad a\in  [a^{\rm match}_{2},a_{\rm freeze}]\ , ~\kappa\ll 1
\eea
where $ a^{\rm match}_1 \in  [0,a_c]$,  $ a^{\rm match}_2 \in [a_c,a_{\rm freeze} ]$. Now the interior point  $a_0 \in [ a^{\rm match}_1,  a^{\rm match}_2]$  is chosen near the freezing point. Expanding close to this point, we find
\be
-Q(a) \underset{a\to a_0 }{=}  d_{0}+d_{1}(a-a_0 )+d_{2}(a-a_0)^2+\ldots
\ee
and the corresponding equations are locally solved in terms of hypergeometric Polynomial Cylinder functions  $D_{\mu}\left[ x  \right]$ with parameters
\bea
x(a) = \ft{ 2 d_{2}( a {-} \,a_0 ){+}d_{1}}{\sqrt{2} d_{2}^{3/4}}\ , \quad
\mu^\pm = \ft{-4 d_{2}^{3/2}\pm (d_{1}^2-4 d_{0} d_{2})}{8 d_{2}^{3/2}} \ . 
\eea
 Finally, the coefficients $c_i$ are determined by matching the local solutions and their first derivatives  at the matching points. 
  }

\end{itemize}
For each inflationary model specified by $H(a)$, we will integrate $\mathfrak{u}_{k}$ using the three methods, and compute the power spectra   $P_S(k)$, $P_T(k)$ using (\ref{eq:Power spectra definition}). 
 To compute the  logarithmic derivative, we repeat the computations  for  $N_{\rm pts}$  choices of the momenta around $k$ and  use the least squares formula
 \be
 {d\log P(k) \over d\log k}={\sum_{i} \log P(k_i) (\log k_i-\overline{\log k} )\over \sum_i (\log k_i-\overline{\log k} )^2 }\ ,
 \ee
with  
\be
\overline{\log k}=\ft{1}{N_{\rm pts}}\sum_i \log k_i \ .
\ee
  We always check the stability of the results under variations  of  the regulator $N_{\rm reg}$ and the number of points $N_{\rm pts}$. This is a necessary procedure in order to be able to make sense of the extracted values for the spectral indices for each model.
      
\subsection{ Starobinsky-like model}

\begin{figure}[t]
\begin{minipage}{0.5\textwidth}
\centering
\includegraphics[width=1\textwidth]{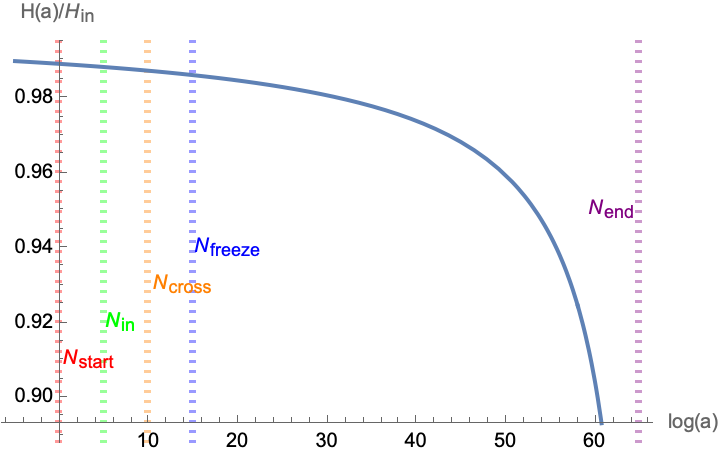}
\end{minipage}
\hfill
\begin{minipage}{0.5\textwidth}
\centering
\includegraphics[width=1\textwidth]{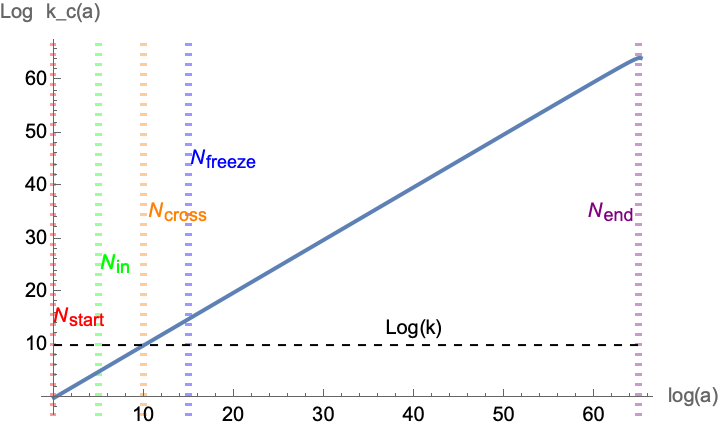}
\end{minipage}
\caption{\it  Starobinsky model:  Hubble function $H(a)/H_{\rm in}$ and crossing momenta $k_c(a)=a H(a)/H_{\rm in}$. }\label{fig_StaroHandk}
\end{figure}

 We first consider the Starobinsky-like model of inflation. We set $\alpha=\sqrt{\frac{2}{3}}$, leading to
\be
H(a)=H_{\rm in}\left(1+\frac{1}{ {\cal W}_{-1}\left(-a^{4\over 3} \right)} \right)\ .
\ee 
 
 \subsection*{Numerical integration}
In figure \ref{fig_StaroHandk} we show  the number of e-foldings $N=\log a$ for the characteristic points $a_{\rm in}$, $a_{\rm cross}$,   $a_{\rm cross}$, $a_{\rm freeze}$ along evolution, for a mode  with momenta
\be
k=21748 \, H_{\rm in}\ , \label{khin}
\ee
 chosen such that $a_c=e^{10}$.  The power spectra obtained from the numerically integrated solution are displayed in figure \ref{fig:matchvsnum sca staro}. 
 Comparing against the experimental observed value
\be
\frac{2}{5}P_S^{1\over 2}=2.1\cdot10^9\ ,
\ee
one can determine the constant
\be
H_{\rm in}=7.817\cdot10^{-7}\ .
\ee
Plugging this into (\ref{khin}) one finds that $k$ matches  the pivot scale known in literature $k_\star=0.05  \,\text{Mpc}^{-1}$.

  Logarithmic derivatives are computed from $N_{\rm pts}=10$ points distributed around $k$ with step $\delta=100 H_{\rm in}$. 
   Setting $N_{\rm reg}=5$, the results for the spectral indices are
\be
n_S=0.9653, \qquad
n_T=-0.0006, \qquad
r=0.0034.
\ee
Varying $N_{\rm reg}$ one finds oscillations on the last digit, so this sets the error scale for our numerical computation.  

{
\subsection*{Semi-analytic integration I}
In order to compute the power spectra we have to choose the points $a_1^{\rm match}$ and $a_2^{\rm match}$:
\bea
\begin{array}{cccc}
                    &  ~~~~\kappa\left(a^{\rm match}_1\right)~~~  & ~~~\kappa( a_c)~~~~ & ~~~~\kappa\left(a^{\rm match}_2\right) ~~~~ \\
 {\rm Hankel\,\,Ansatz } &       3 &  1 &  0.01\\
\end{array}
\eea
 We use the same choice for both scalar and tensor perturbations. Solving the matching conditions one finds
\begin{equation}
\label{tab:ci staro Hankel}
\small % Mantiene la dimensione del testo proporzionata senza sgranare i font
\setlength{\extrarowheight}{4pt} % Alternativa pulita a \arraystretch
\begin{array}{ccccc}
 & c_1 & c_2 & c_3 & c_4  \\
\text{Scalar1H} & {-}0.89{-}0.02 {\rm i} ~~~& {-}(4+1.4 {\rm i} )\times 10^{{-}5}~~~ & (1.6{-}87 {\rm i} )\times 10^{6}~~~ & {-}1.9\times 10^{-7}{+}10^{-5} {\rm i} \\
\text{Tensor1H} & {-}0.89 & {-}(5.9{+}91 {\rm i} )\times 10^{{-}7} & (9.92{-}9.69i)\times 10^{7} & (2.55{+}22187i)\times 10^{{-}11} \\
\end{array}
\end{equation}
We show in figure \ref{fig:matchvsnum sca staro} the power spectrum and  in table \ref{tab:parameters long matching} the results for the spectral indices. 
}

\subsection*{Semi-analytic integration II}

   We notice that the choice of the points $a_0$ and $a_i^{\rm match}$ affects  the deviations of the piecewise function from the actual numerical solution, so we choose them in order to minimize the error. 
    The following choice turns out to be optimal at approximating the Starobinsky-like model:
 \bea
\begin{array}{cccc}
                    &  ~~~~\kappa\left(a^{\rm match}_1\right)~~~  & ~~~\kappa( a_0)~~~~ & ~~~~\kappa\left(a^{\rm match}_2\right) ~~~~ \\
 {\rm Scalar 1D } &       3 &  0.1833 &  0.1705\\
  {\rm Tensor 1D} &       3 &  0.1778 &  0.1500\\
\end{array}
\eea
{
 leading to 
\begin{equation}
\label{tab:ci staro 1D}
\small % Mantiene la dimensione del testo proporzionata senza sgranare i font
\setlength{\extrarowheight}{4pt} % Alternativa pulita a \arraystretch
\begin{array}{ccccc}
 & c_1 & c_2 & c_3 & c_4  \\
\text{Scalar1D} & 4610{-}1338i~~~ & {-}8164{+}5922i ~~~& 621623{-}834177i ~~~& ({-}5.89{+}8.71i)\times 10^{{-}6} \\
\text{Tensor1D} & 4688{-}1312i & {-}8268{+}5950i & (37.6{-}5.59i)\times10^{6} & ({-}1.21{+}1.80i)\times10^{6}  \\
\end{array}
\end{equation}
}

\begin{figure}[t]
    \centering
    \includegraphics[width=0.49\linewidth]{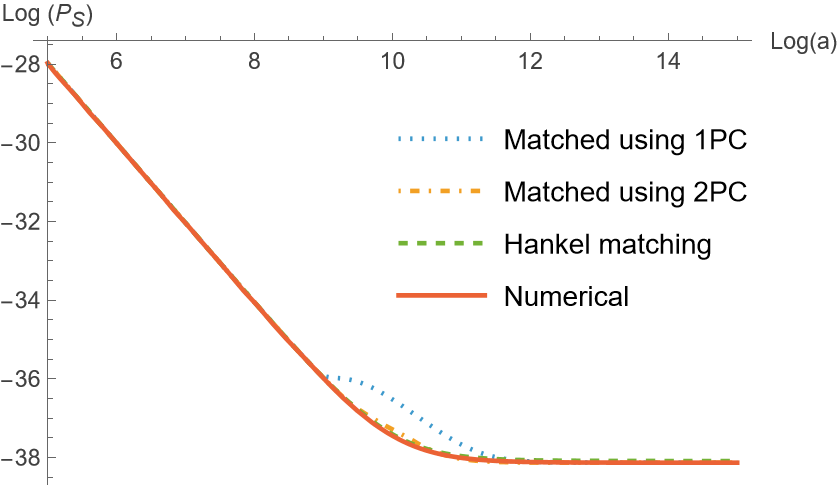}
      \includegraphics[width=0.49\linewidth]{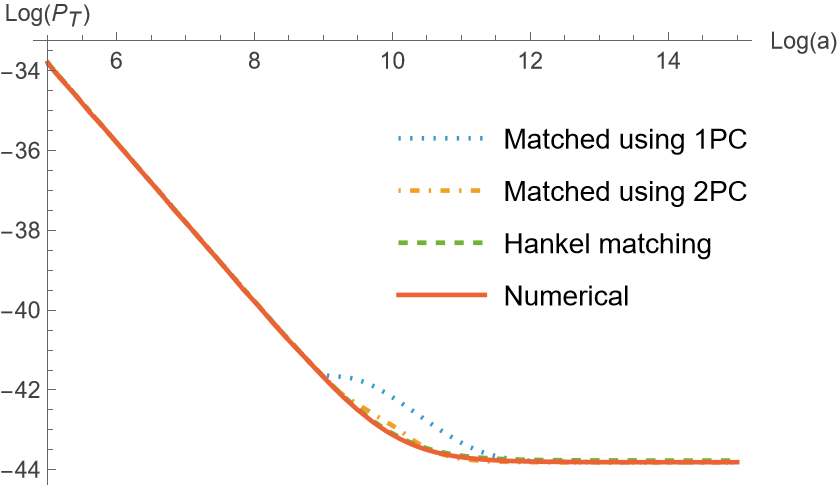}
    \caption{\it Matched vs Numerical scalar and tensor power spectra for the Starobinski model.}
    \label{fig:matchvsnum sca staro}
\end{figure}

 We display the resulting power spectrum in figure \ref{fig:matchvsnum sca staro}. We notice that the semi-analytic solution  perfectly matches the numerical one in the region of interest, \emph{i.e.} where freezing takes place. This is where the spectral indices are computed, hence it will produce 
 excellent results for the spectral indices, although our solution actually still deviates from the numerical solution in the intermediate region.   The matching in this intermediate region can be further improved by adding an extra interior joining point, where we locally solve the equations again by
 using Polynomial cylinder functions.  In  figure \ref{fig:matchvsnum sca staro} we display the results for this choice.
 \bea
\begin{array}{cccccc}
                    &  ~~~~\kappa\left(a^{\rm match}_1\right)~~~  & ~~~\kappa( a_0)~~~~ & ~~~~\kappa\left(a^{\rm match}_2\right) ~~~~ &~~~~\kappa\left(a_1\right) ~~~~ &~~~~\kappa\left(a^{\rm match}_3\right) ~~~~  \\
 {\rm Scalar 2D } & 2.0 &  1.15 &  0.74& 0.48& 0.45\\
  {\rm Tensor 2D} & 2.3 &  1.29 &  0.85 & 0.61 & 0.41\\
\end{array}
\eea
{
%We show the coefficients obtained in table \ref{tab:ci staro 2D}.
%\begin{equation}
%\label{tab:ci staro 2D}
%\resizebox{\textwidth}{!}{$%
%\setlength{\extrarowheight}{4pt}
%\begin{array}{|c|c|c|c|c|c|c|}
%\hline
% & c_1 & c_2 & c_3 & c_4 & c_5 & c_6 \\
%\hline
%\text{Scalar2D} & 18.5-23.7i & -46.4+141.6i & 492.9-595i & -618.7+1546i & 56462-150884i & (-6+105i)\times10^{-7} \\
%\hline
%\text{Tensor2D} & 22.8-12.3i & -59.3+117.7i & 356.4-434.6i & -480+1136i & (4.8-1.7i)\times10^{6} & (-2.09+21.6i)\times10^{-8} \\
%\hline
%\end{array}$%
%}
%\end{equation}
}

Table \ref{tab:parameters long matching} reports the comparison of the results against those obtained by via numerical and slow-roll approximation.   
We observe that both the numerical, semi-analytical and slow-roll results are compatible with the observational constraints, telling us that our model is a viable one for inflation. \\

\begin{table}[h]
    \centering
    \begin{tabular}{|c|c|c|c|}
         \hline
         \textbf{Method} & $n_s$ & $n_T$ & $r$ \\
        \hline
        Observational constraints & $0.9649\pm0.0042$ & $-1.37<n_T<0.42$ & $r<0.028$ \\
        \hline
        Slow-roll approximation & 0.962 & -0.00052 & 0.00413 \\
        \hline
        Numerical & $0.9653$ & -0.0006 & 0.0034 \\
        \hline
        Hankel approach & 0.9659 & 0.00022 & 0.0034 \\
        \hline
        Semi-analytical 1PC & $0.9647$  & -0.00057   & 0.0034 \\
        Semi-analytical 2PC & 0.9655 & -0.00054 & 0.0034\\
        \hline
    \end{tabular}
    \caption{\it Spectral indices of the Starobinsky-like model.}
\label{tab:parameters long matching}
\end{table}

\subsection{T-model}

The same results can be obtained for other models in our list. Here we give some details for the T-model. 
We set  $\alpha=\frac{1}{4}$ and $n=2$, leading to the Hubble function  
\be
H(a)=H_{\rm in}\frac{\log(a)+1}{\log(a)-1} \quad .
\ee
 We start from a mode with pivot scale
\be
k=21263 H_{\rm in} \ ,
\ee
 crossing the Hubble horizon after 10 e-folds. Comparing against the experimental value for $P_S$ we find that   
\be
H_{\rm in}=1.134\times 10^{-6}\ .
\ee
\begin{figure}
    \centering
    \includegraphics[width=0.49\linewidth]{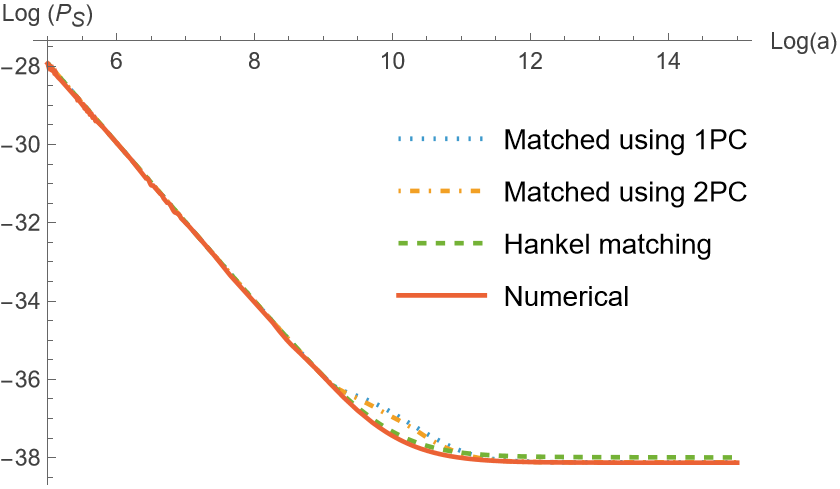}
      \includegraphics[width=0.49\linewidth]{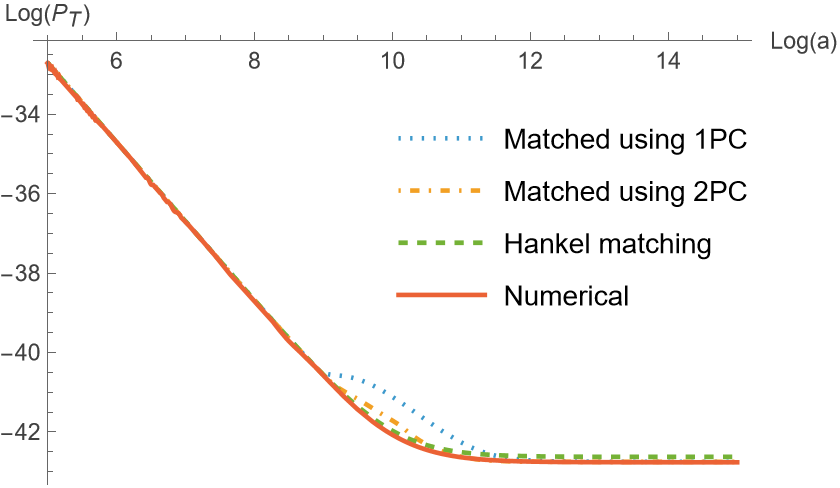}
    \caption{\it Matched vs Numerical scalar and tensor power spectra for the T-model.}
    \label{fig:matchvsnumT}
\end{figure}
{
\subsection*{Semi-analytic integration I}
Now we choose the matching points:
\bea
\begin{array}{cccc}
                    &  ~~~~\kappa\left(a^{\rm match}_1\right)~~~  & ~~~\kappa( a_c)~~~~ & ~~~~\kappa\left(a^{\rm match}_2\right) ~~~~ \\
 {\rm Hankel\,\,Ansatz } &       3 &  1 &  0.01\\
\end{array}
\eea
leading to
\begin{equation}
\label{tab:ci Tmod Hankel}
\small % Mantiene la dimensione del testo proporzionata senza sgranare i font
\setlength{\extrarowheight}{4pt} % Alternativa pulita a \arraystretch
\begin{array}{ccccc}
 & c_1 & c_2 & c_3 & c_4  \\
\text{Scalar1H} & -0.886-0.016i~~~ & -(-2-15i)\times 10^{-4}~~~ & (2.7-152.1i)\times 10^{6} ~~~& (-1.2+66.6i)\times 10^{-7}\\
\text{Tensor1H} & -0.886-0.0003i & -(5.1+26.6i)\times 10^{-6} & (2.3-101.6i)\times 10^{23} & (7.4+23207.9i)\times 10^{-11} \\
\end{array}
\end{equation}
    We show in figure \ref{fig:matchvsnumT}  and spectral indices  in table \ref{tab:parameters long matching Tmod}. 
    }
 
\subsection*{Semi-analytic integration II}
The semi-analytic computation with one internal point makes the choice
 \bea
\begin{array}{cccc}
                    &  ~~~~\kappa\left(a^{\rm match}_1\right)~~~  & ~~~\kappa( a_0)~~~~ & ~~~~\kappa\left(a^{\rm match}_2\right) ~~~~ \\
 {\rm Scalar 1D } &       2.5 &  0.2285 &  0.183\\
  {\rm Tensor 1D} &       3.0 &  0.1778 &  0.150\\
\end{array}
\eea
{
We show the coefficients obtained in table \ref{tab:ci Tmod 1D}.
\begin{equation}
\label{tab:ci Tmod 1D}
\small % Mantiene la dimensione del testo proporzionata senza sgranare i font
\setlength{\extrarowheight}{4pt} % Alternativa pulita a \arraystretch
\begin{array}{ccccc}
 & c_1 & c_2 & c_3 & c_4  \\
\text{Scalar1D} & 2505{-}1788i ~~~ & {-}3701{+}5310i ~~~& 312571{-}856253i ~~~& ({-}1.62{+}6.04i)\times 10^{-6} \\
\text{Tensor1D} & 4591{-}1516i & {-}7930{+}6283i & (3.01{-}4.48i)\times10^{-22} & ({-}1.13{+}1.86i)\times10^{-7}  \\
\end{array}
\end{equation}
}
while with two internal point, we take
 \bea
\begin{array}{cccccc}
                    &  ~~~~\kappa\left(a^{\rm match}_1\right)~~~  & ~~~\kappa( a_1)~~~~ & ~~~~\kappa\left(a^{\rm match}_2\right) ~~~~ &~~~~\kappa\left(a_2\right) ~~~~ &~~~~\kappa\left(a^{\rm match}_3\right) ~~~~  \\
 {\rm Scalar 2D } & 2.5 &  1.15 &  0.74& 0.2701& 0.25\\
  {\rm Tensor 2D} & 2.5 &  1.29 &  0.84 & 0.718 & 0.41\\
\end{array}
\eea
{
%We show the coefficients obtained in table \ref{tab:ci Tmod 2D}.
%\begin{equation}
%\label{tab:ci Tmod 2D}
%\resizebox{\textwidth}{!}{$%
%\setlength{\extrarowheight}{4pt}
%\begin{array}{|c|c|c|c|c|c|c|}
%\hline
% & c_1 & c_2 & c_3 & c_4 & c_5 & c_6 \\
%\hline
%\text{Scalar2D} & 31.8-21.8i & -79.1+154.6i & 1735-1378i & -2483+3959i & 217757-619060i & (-1.34+6.11i)\times10^{-6} \\
%\hline
%\text{Tensor2D} & 26.8-10.3i & -68.6+120.5i & 671.3-361.8i & -859+1220i & (3.8-9.1i)\times10^{-23} & (-3.2+21.6i)\times10^{-8} \\
%\hline
%\end{array}$%
%}
%\end{equation}
}
 The results for the power spectrum are displayed in figure \ref{fig:matchvsnumT},  while spectral indices are reported in table \ref{tab:parameters long matching Tmod}.  \\

 \begin{table}[h]
    \centering
    \begin{tabular}{|c|c|c|c|}
         \hline
         \textbf{Method} & $n_s$ & $n_T$ & $r$ \\
        \hline
        Observational constraints & $0.9649\pm0.0042$ & $-1.37<n_T<0.42$ & $r<0.028$ \\
        \hline
        Slow-roll approximation & $0.9635$ & $-0.00124$ & $0.0099$ \\
        \hline
        Numerical & $0.9627$ & $-0.00254$ & $ 0.0096$ \\
        \hline
        Hankel approach & $0.9656$ & $0.00061$ & $0.0097$ \\
        \hline
        Semi-analytical 1PC & $0.9629$  & $-0.00167$   & $0.0096$ \\
        Semi-analytical 2PC & $0.9637$ & $-0.00166$ & $ 0.0096$ \\
        \hline
    \end{tabular}
    \caption{\it Spectral indices and tensor-to-scalar ratio of the T-model.}
\label{tab:parameters long matching Tmod}
\end{table}

\section{Conclusions and Outlook}

Let us conclude and summarize our results. We have identified a large class of integrable cosmological models based on a single self-interacting scalar field minimally coupled to gravity that mimics a general multi-component fluid, thus providing a joint description of the different phases of the universe from inflation to kination, from radiation to (dark) matter and more\footnote{Due its mimetic nature, it is tempting to call 'our' scalar field a `chameleon'  alas this appellative has already been used in cosmology for a different purpose \cite{Khoury:2003aq, Khoury:2003rn}.}. The approach heavily relies on the observation that the Friedman equation relates the scalar potential $V(\phi)$  to the Hubble function $H(\phi)$  via (\ref{vh}), the latter playing the role  of the fake superpotential just as the one introduced in the context of  supergravity domain-walls \cite{Skenderis:2006jq}.

  We have constructed several integrable models that closely resemble popular inflationary models, based on Starobinski, $\alpha$-attractors, polynomial, hyperbolic and trigonometric potentials. The models obtained in this way admit by construction a first order formulation, which admits analytic solutions  for the inflationary background for any given $H(\phi)$, upon choosing the appropriate time coordinate. We study scalar and tensor perturbations of these backgrounds, using the scale factor $a$ as time coordinate following the lines of \cite{Bianchi:2024mlq}. Bearing this in mind, out of all possibile models, we focus on those where the dependence $a(\phi)$ can be explicitly integrated and inverted into a $\phi(a)$.  
The virtue of our approach is the possibility of studying cosmological perturbations beyond the slow-roll approximation. In this framework,  we integrated numerically and semi-analytically (i.e. piecewise analitically) the linearized equations governing both scalar and tensor perturbations and extracted the relevant observables: the spectral indices $n_S$ and $n_T$, the power spectra and the tensor-to-scalar ratio $r$.  We compared the results against experimental data and predictions based on the slow-roll approximation and find excellent agreement.  

The natural next step would be to extend this analysis to hot big bang cosmologies, along the lines of \cite{Bianchi:2024mlq}., where  quantum Seiberg-Witten techniques can be exploited to connect different eras of the cosmological evolution. 
Most importantly one would like to address interesting cosmological phases such as reheating after inflation in mode details or else consider (first order) phase transitions that can lead to bubble nucleation and (primordial) GW production. 
Another interesting line of research would be to investigate multi-field generalizations, along the lines of  \cite{Skenderis:2006jq, McFadden:2009fg}. Within this (holographic) framework \cite{McFadden:2010vh} one can study and estimate non-gaussianities \cite{Bartolo:2004if, Babich:2004gb, Maldacena:2011nz} or higher-point correlators that can be tackle within the so-called 'cosmological  boot-strap' approach \cite{Arkani-Hamed:2018kmz, Baumann:2020dch, Sleight:2019hfp} or via `analytic continuation' from AdS (to dS) \cite{Sleight:2020obc, Sleight:2021plv}. 

Last but not least one would like to cook up models with an intermediate plateau that can efficiently trigger the production of primordial BHs (PBHs) in interesting mass ranges for them to play a role as dark matter \cite{Franciolini:2018vbk} that represent one of the goals of LISA  \cite{LISACosmologyWorkingGroup:2023njw}.

\section*{Acknowledgements}
We would like to especially thank A.~Riotto, P.~Creminelli, F.~Arcofora, F.~Fucito and A.~Marcian\`o  for useful discussions. 
M.~B., G.~D. and J.~F.~M. thank the MIUR PRIN contract 2020KR4KN2 \lq\lq String Theory as a bridge
between Gauge Theories and Quantum Gravity'' and the INFN project ST\&FI \lq\lq String Theory and
Fundamental Interactions'' for partial support.

\bibliographystyle{JHEP}
\bibliography{ref}

\end{document}